\documentclass[11pt,a4paper,oneside]{article} 
\usepackage[T1]{fontenc} % New encoding (with fix-cm, the look & feel of
\usepackage{lmodern}
\usepackage{microtype} % character protrusion, font expansion, etc.
\usepackage[pdftex]{graphicx}
\usepackage[usenames,dvipsnames]{color}
\usepackage{amsmath,amsfonts,amssymb}
\usepackage[small]{titlesec}
\usepackage[english]{babel}
\usepackage[sectionbib]{natbib}
\usepackage{enumitem}
\usepackage[font=small,labelfont=bf,labelsep=period]{caption}
\usepackage[pdftex,hyperfigures,breaklinks,colorlinks]{hyperref}
\usepackage[hmargin=2.5cm,vmargin={2.0cm,2.0cm}]{geometry}
\usepackage[raggedright]{sidecap}
\usepackage{fixltx2e}

\definecolor{LinkColor}{rgb}{0, 0, 0.3}
\definecolor{ExtLinkColor}{rgb}{0, 0.3, 0}
\hypersetup{citecolor=LinkColor,linkcolor=LinkColor,urlcolor=ExtLinkColor}

\sidecaptionvpos{figure}{c}
\newcommand{\ee}{\mathbb{E}}
\begin{document}
\title{Cavity Method: Message Passing from a Physics Perspective}
\date{}
\author{Gino Del Ferraro, KTH Royal Institute of Technology, Stockholm \\
Chuang Wang, Institute of Theoretical Physics, Chinese Academy of
Sciences, China\\
Dani Mart\'{i}, \'{E}cole Normale Sup\'{e}rieure \& Inserm, France\\
Marc M\'{e}zard, Universit\'{e} Paris-Sud \& CNRS, France}

\maketitle

\begin{quotation}
  \itshape
  \noindent
  \small
These are the notes from the lecture by Marc M\'{e}zard given at the autumn school ``Statistical Physics, Optimization, Inference, and Message-Passing Algorithms'', which took place at Les Houches, France, from September 30th to October 11th 2013. The school was organized by Florent Krzakala from UPMC \& ENS Paris, Federico Ricci-Tersenghi from La Sapienza Roma, Lenka Zdeborova from CEA Saclay \& CNRS, and Riccardo Zecchina from Politecnico Torino.
\end{quotation}

\abstract{In this three-sections lecture cavity method is introduced as heuristic framework from a Physics perspective to solve probabilistic graphical models and it is presented both at the replica symmetric (RS) and 1-step replica symmetry breaking (1RSB) level. This technique has been applied with success on a wide range of models and problems such as spin glasses, random constrain satisfaction problems (rCSP), error correcting codes etc. Firstly, the RS cavity solution for Sherrington-Kirkpatrick model---a fully connected spin glass model---is derived and its equivalence to the RS solution obtained using replicas is discussed. Then, the general cavity method for diluted graphs is illustrated both at RS and 1RSB level. The latter was a significant breakthrough in the last decade and has direct applications to rCSP. Finally, as example of an actual problem, K-SAT is investigated using belief and survey propagation.}

\tableofcontents

\label{Chap: Mezard}

%\begin{center}
%  \LARGE \scshape Marc M\'{e}zard
%\end{center}

%\begin{center}
%\includegraphics[width=0.6\textwidth]{lecturer_pic}
%\end{center}

%%\begin{center}
%\noindent Marc M\'{e}zard\\
%Universit\'{e} Paris-Sud \& CNRS, LPTMS,\\
%UMR8626, B\^at. 100,\\
%91405 Orsay, France.
%\begin{center}
%\emph{Notes written in collaboration with} \\
%Gino Del Ferraro, KTH Royal Institute of Technology, Stockholm \\
%Chuang Wang, Institute of Theoretical Physics, Chinese Academy of
%Sciences, China\\
%Dani Mart\'{i}, \'{E}cole Normale Sup\'{e}rieure \& Inserm, Paris,
%France\\
%\end{center}

\newpage

%%%%%%%%%%%%%%%%%%%%%%%%%%%%%%%%%%%%%%
\section{Replica solution without replicas}
\label{sec:one}
\subsection{The Sherrington-Kirkpatrick model}
The Sherrington Kirkpatrick (SK) model \citep{sk1975prl} is a mean-field version of the Edward-Anderson Model \citep{edwards1975theory} and it is defined by a system of $N$ Ising spins $\sigma = (\sigma_1, \sigma_2,\dotsc,\sigma_N)$ taking values $ \pm 1$ placed on the vertices of a lattice. In the SK mean field description the model is fully connected: every spin interacts with everybody else, and the couplings $J_{ij}$ are chosen independent and identically distributed according to a gaussian probability distribution, such that, the probability distribution of the whole couplings reads
\begin{equation*}
  P(J) = \prod_{i <j}P(J_{ij}) \propto \exp\left(-\frac{N}{2} \sum_{i <j} J^2_{ij}\right).
\end{equation*}
The  $J_{ij}$ variables are assumed to be symmetric and not having self interacting terms, i.e., $J_{ij}=J_{ji}$ and $J_{ii}=0$, we stress here that physically they play the role of quenched disorder among each couple of spin in the system. By quenched disorder we mean that the couplings $J$ exert a stochastic external influence on the system, but they don't participate to the thermal equilibrium. The Hamiltonian of the system, given a particular configuration $\sigma$, is given by
\begin{equation*}
  H_J(\sigma) = - \sum_{i<j} J_{ij}\sigma_i \sigma_j - h \sum_i \sigma_i,
\end{equation*}
where $h$ is the homogeneous external magnetic field on each site $i$, and the couplings $J_{ij}$ are of the order of $1/\sqrt{N}$ to ensure a correct thermodynamic behaviour of the free energy. In this lecture we will be interested in equilibrium properties of the system; the probability distribution at equilibrium is then given by the Boltzmann-Gibbs distribution,
\begin{equation*}
  P(\sigma) = \frac{1}{Z} \exp\bigl(-\beta H_J(\sigma)\bigr),
\end{equation*}
where we introduced the \emph{partition function},
\begin{equation*}
  Z = \sum_{\sigma} \exp\bigl(-\beta H_J(\sigma)\bigr),
\end{equation*}
which includes a sum over all the possible spin configurations, which we denote by $\{\sigma\}$.

\begin{figure}[htbp]
  \centering
  \includegraphics{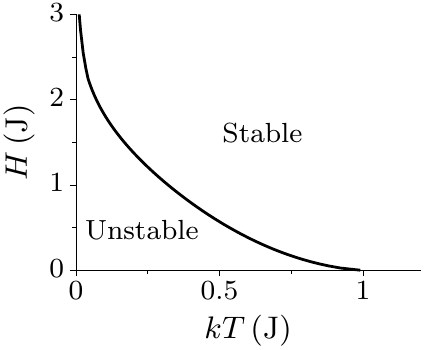}
  \caption{Phase diagram showing the limit of stability of the Sherrington-Kirkpatrick solution for the paramagnetic phase in the presence of a magnetic field $h$.}
  \label{fig:AT_line}
\end{figure}

\begin{figure}[htbp]
  \centering
  \includegraphics{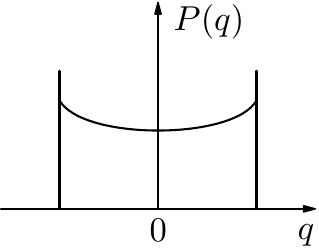}
  \caption{Distribution function $P(q)$ of the SK model with a Full RSB ansatz, \emph{i.e.} a system with multi valley structure.}
  \label{fig:Pq}
\end{figure}

The phase diagram $h$ vs.\ $T$ for this problem, relative to the stability of the replica symmetric (RS) solution, was found by de Almeida and Thouless \citep{de1978stability} and is shown in Figure~\ref{fig:AT_line}.
We observe that there are two phases: in the high temperature regime there is a paramagnetic phase and in the low temperature regime there is a spin-glass phase where the RS solution is unstable. The transition line between these two phases is called the de Almeida-Thouless line. We can then define an order parameter that allows us to distinguish between these two phases. Let us consider two copies of the same system, which are two different spin configurations $\sigma$ and $\tau$ with associated probability $P(\sigma)$ and  $P(\tau)$. Then, defining the overlap between these two configurations as $q_{\sigma \tau} = \frac 1 N \sum_{i} \sigma_i \tau_i$, it is possible to compute the probability that this overlap is equal to $q$ as follows,
\begin{equation*}
P_J(q)= \lim_{N \to \infty} \sum_{\sigma \tau} P_J(\sigma)P_J(\tau) \delta(q_{\sigma\tau}-q),
\end{equation*}
In principle the probability of having a given overlap configuration depends on the sample, i.e., on the disorder, which means that we need to take the average over the disorder to remove this dependence, namely $P(q)=  \mathbb{E}_J P_J(q)$, where $\mathbb{E}_J$ is the average over the disorder. The probability distribution $P(q)$ in the case of Replica Symmetry Breaking (RSB) ansatz, is shown in Figure~\ref{fig:Pq}

\subsubsection{Pure states} 
The RSB solution of the Sherrington-Kirkpatrick (SK) model is characterized by the order parameter matrix $Q_{ab}$ (as shown by G. Parisi in his lectures). Since this system presents spontaneous symmetry breaking, if there is a particular solution for the matrix $\hat Q$ with the RSB, then any other matrix obtained via any permutation of the replica indices in $\hat Q$ will also be a solution. On the other hand, within the mean field approximation, because the total free energy is proportional to the volume of the system, the energy barriers separating the corresponding ground states must be infinite in the thermodynamics limit. As a consequence, once the system is found to be in one of these states, it will never be able to jump into another one in a finite time. In this sense, the observable state is not the Gibbs one, but one of these states. To distinguish them from the Gibbs states, they could be called \emph{pure states} and the probability measure can be decomposed as the sum of the measures over the pure states. According to this definition, the average of any observable $\mathcal{O}$ can be taken as the sum of the averages in each of the pure states, as follows:
\begin{equation*}
  \langle \mathcal{O} \rangle = \sum_\alpha w_\alpha \mathcal{O} , \quad \text{with} \quad  w_\alpha= \frac{\mathrm{e}^{-\beta F_\alpha}}{\sum_\alpha \mathrm{e}^{-\beta F_\alpha}},
\end{equation*}
where $F_\alpha$ is the free energy associated to the pure state $\alpha$. More formally, the pure states could be defined as those in which the correlation function of two spin variables belonging to the same pure states tends to zero in the thermodynamic limit, i.e., $\langle \sigma_i \sigma_j \rangle_\alpha -\langle \sigma_i \rangle_\alpha \langle \sigma_j \rangle_\alpha \to 0$ as $N \to \infty$.

\subsubsection{The Cavity Method in the RS case}
We now investigate an alternative method with respect to the replica trick used so far to investigate the SK model from which is possible to recover all the results at the RS level \citep{mezard1986sk}. This method can be also viewed as an analytic ansatz to derive and analyze the Thouless-Anderson-Palmer (TAP) equations~\citep{thouless1977solution}. The basic idea is to go from an SK system $\Sigma_N$ composed of $N$ spins to a $\Sigma_{N+1}$ system that has $N+1$ spins, assuming that the thermodynamic limit exists, or in other words, assuming that in the thermodynamic limit there is no difference between observables computed in both systems (as for instance the free energy). We shall make some physical assumption on the organisation of the configuration of $\Sigma_N$ inspired from the results obtained in the SK model with the RSB ansatz by using replicas \citep{parisi1979infinite,parisi1980order}: the ultrametric organisation of the states and the independent exponential distribution of their free energies. Once this properties are assumed to be valid in $\Sigma_N$ we will show that they are valid also for $\Sigma_{N+1}$ and so, for instance
$\overline{\langle \sigma_i \rangle^2_N}=\overline{\langle \sigma_i \rangle^2_{N+1}}$ as $N \to \infty$, where the bar denotes average over $J$. Let's assume that $\sigma_0$ is the spin added to the system of $N$ spins to create the $N+1$ spins system. The probability distributions of disorder in each of them are respectively
\begin{align*}
  P_N(J) &= \prod_i P_N(J_{ij}) \propto \exp\biggl(-\frac{N}{2} \sum_{i <j} J^2_{ij}\biggr), \\ 
   P_N(J,J_0) &=\prod_{j,i<j} P_{N+1}(J_{ij}, J_{0j}) = \prod_{j} P_{N+1}(J_{0j}) \prod_{i<j}P_{N+1}(J_{ij}) \\ 
   & \propto \exp\biggl(-\frac{N+1}{2} \biggl[\sum_{i <j} J^2_{ij} +\sum_j J_{0j}\biggr]\biggr),
\end{align*}
where $J_{0j}$ is the coupling between the added spin $\sigma_0$ and all the other spins in the $\Sigma_{N+1}$ system and we also note that there is a small change scale of $J$ (from $N\to N+1$ in the exponent). Then the probability distribution of a certain configuration of spin in the $N+1$ system is given by:
\begin{equation*}
% P_N(\{\sigma_i\},\sigma_0)=\exp\Bigl(-\beta H_N(\{\sigma_i\})+\beta \sum_j^c J_{0j}^2 \sigma_0\sigma_j\Bigr),
P_N(\sigma, \sigma_0)=\exp\Bigl(-\beta H_N(\sigma)+\beta \sum_j  J_{0j} \sigma_0\sigma_j\Bigr),
\end{equation*}
where $\sigma=(\sigma_1, \dotsc, \sigma_N)$, and $h^c \equiv \sum_j J_{0j}\sigma_j$ is the local field felt by all the other spins in the $\Sigma_{N+1}$ system because of the presence of $\sigma_0$. The index $c$ indicates the ``cavity'', since $h^c$ is usually called ``cavity field''. In the following we want to compute the probability distribution of $h^c$. To do this, we will compute all the moments of the distribution. Let's start by defining the \emph{non-linear susceptibility} as
\begin{equation}
  \chi= \frac{1}{N} \sum_{i<j} \bigl(\langle \sigma_i \sigma_j \rangle_ -\langle \sigma_i \rangle \langle \sigma_j \rangle \bigr)^2,
\label{eq:chi}
\end{equation}
and computing the expectation and variance of the cavity field:
\begin{align}
  \bigl\langle h^c \bigr\rangle_N  &=  \sum_i J_{0i} \langle \sigma_i \rangle_N \xrightarrow{N \to \infty} h  \label{eq:h1}\\
\bigl\langle (h^c)^2 \bigr\rangle_N - \bigl\langle h^c \bigr\rangle^2_N  &=  \sum_{i,j} J_{0i}J_{0j} \bigl(\langle \sigma_i\sigma_j\rangle_N - \langle \sigma_i \rangle_N \langle \sigma_j \rangle_N\bigr)^2 \label{eq:h2}
\end{align}
The assumption of the cavity method at the RS level is that the susceptibility \eqref{eq:chi} has to be finite. Because $J_{ij}$ is of the order of $1/\sqrt{N}$, and because the sum over $i,j$ involves $N^2$ terms, $\chi$ will be finite as long as the connected correlation of $\sigma_i$ and $\sigma_j$, namely $\langle \sigma_i\sigma_j\rangle_N - \langle \sigma_i \rangle_N \langle \sigma_j \rangle_N$, is of order $1/\sqrt{N}$. Then if we take the sum in \eqref{eq:h2} will be dominated by the term $i=j$:
\begin{equation*}
  \text{for $i=j$:} \quad  \bigl\langle (h^c)^2 \bigr\rangle_N - \bigl\langle h^c \bigr\rangle^2_N = \sum_i J_{0i}^2 (1-\langle \sigma_ i \rangle^2_N)= 1 - \frac 1 N \sum_i \langle \sigma_i \rangle^2 =1- \overline{\langle \sigma_i \rangle^2} = 1-q, 
\end{equation*}
where in the second equality we used $J_{ij} \sim 1/\sqrt{N}$, while in the third we substituted the sum over all sites with the average over the disorder at a single site, because they are equivalent. Finally we used the definition of the Edwards-Anderson order parameter $\overline{\langle \sigma_i \rangle^2} = q$. Using similar reasonings, one can compute the forth moment, 
\begin{equation*}
\Bigl\langle (h^c- \langle h^c\rangle)^4 \Bigr\rangle= \sum_{i,j,k,l} J_{0i}J_{0j} J_{0k}J_{0l} \langle (\sigma_i - \langle \sigma_i \rangle)(\sigma_j - \langle \sigma_j \rangle)(\sigma_k - \langle \sigma_k \rangle)(\sigma_l - \langle \sigma_l \rangle) \rangle = 3(1-q)^2,
\end{equation*}
Iterating this computation and applying similar considerations, we claim that all odd moments bigger than the first one are zero, while all even moments are given by the following expression:
\begin{equation}
\Bigl\langle(h^c)^{2p} \Bigr\rangle = (2p-1)!!(1-q)^p .
\label{eq:hc}
\end{equation}
These are the moments of a Gaussian distribution with variance $(1-q)$, and therefore the probability distribution of the cavity field in the $\Sigma_N$ systems is given by
\begin{equation}\label{eq:P_n}
P_N(h^c) \cong \exp\left(- \frac{(h^c-h)^2}{2(1-q)}\right),
\end{equation}
where $\cong$ means `equal up to a normalization constant', and $h= \sum_i J_{0i} \langle \sigma_i \rangle_N = \langle h^c \rangle$ is the average value of the cavity field. We stress that the only assumption taken so far in computing these moments has been that the connected correlation function is of order $1/\sqrt{N}$. Now we can consider the probability distribution of $h^c$ in the $\Sigma_{N+1}$ system, which is build by adding the spin $\sigma_0$ to the system $\Sigma_N$:
\begin{equation}
P_{N+1}(h^c,\sigma_0) \cong \exp\left(- \frac{(h^c-h)^2}{2(1-q)} + \beta \sigma_0 h^c\right).
\label{eq:Pn1}
\end{equation}
With this joint distribution it is finally possible to compute many things, like, e.g., the expectation value of the spin $\sigma_0$ in the $\Sigma_{N+1}$ system
\begin{equation}
\langle \sigma_0 \rangle_{N+1} = \tanh(\beta h) = \tanh
\bigl(\beta \sum_{i=1}^{N} J_{0i} \langle \sigma_i \rangle_N\bigr)
\label{eq:s0}
\end{equation}
where the average is taken with respect to the probability density \eqref{eq:Pn1}, integrating over the cavity field $h^c$. This is one of the first results where there is an evident connection, a mathematical relation, between the system $\Sigma_{N+1}$ and the system $\Sigma_{N}$.
Let's then compute the order parameter $q$ from its definition, by using the probability density in Eq.~\eqref{eq:Pn1},
\begin{equation}
  q=\overline{\langle \sigma_0 \rangle^2_{N+1}} = \overline{\tanh(\beta h)^2} 
\end{equation}
where the second equality comes from Eq.~\eqref{eq:s0}.
To compute this average, we need to derive the probability distribution of the cavity field, $P(h)$.  Let us compute its moments. The averaged field reads:
\begin{equation}
  \overline{h}= \sum_i \overline{ J_{0i} \langle \sigma_{0i}\rangle_N}=0,
\end{equation}
which is equal to zero because the average of the couplings $J$'s is zero. The average squared field reads
\begin{equation}
  \overline{h^2}= \sum_{i,j} \overline{J_{0i}J_{0j} \langle \sigma_{0i}\rangle_N \langle \sigma_{0j}\rangle_N} = 
\begin{cases}
  \displaystyle\frac{1}{N} \sum_i \langle \sigma_i \rangle^2_N = q & \quad \text{if $i=j$}, \\
  0 &\quad \text{if $i\neq j$}.
\end{cases}
\end{equation}
By computing all the higher-order moments, it is possible to show that all the odd moments are zero, while all the even ones obey a similar relation to that seen in Eq.~\eqref{eq:hc}. We can thus conclude that $h$ is Gaussian distributed.

Therefore we get:
\begin{equation}
  q= \int \frac{\rm{d h}}{\sqrt{2 \pi q}} \exp\left(-\frac{h^2}{2q}\right) \tanh(\beta h)^2
\end{equation}

The above equation is the self-consistent equation for the $q$ order parameter, originally found by Sherrington and Kirkpatrick~\citep{sk1975prl}. This equation tells us that there is a phase transition at temperature $T=1$, but this solution is unfortunately wrong. This can be shown looking at the thermodynamics; in particular it is possible to show that the entropy of the system, computed with this method and under its assumption, is negative, which is unphysical. This inconsistence arises because the approach followed is equivalent to the RS assumption when one uses replicas, which is not a right ansatz to solve the model. 

We now go back to the initial assumptions that the susceptibility is finite in the thermodynamic limit. To check the validity of this assumption, we will compute $\chi$ in a system $\Sigma_{N+2}$ composed of $N+2$ spins, and we will check the region where the assumption is valid, or more precisely, the region where $\chi$ remains finite. Since we deal with a $\Sigma_{N+2}$ system, we will have to deal with two cavity fields. The  probability measure in this system reads:
\begin{equation*}
P_{N+2}(\sigma_0, \sigma_{0'}, \sigma) \cong \exp\left(-\beta H_N(\sigma)+\beta h^c \sigma_0 + \beta h^{c'}\sigma_{0'}+\beta J_{00'} \sigma_{0}\sigma_{0'}\right)
\end{equation*}
where $h^{c}= \sum_i J_{0 i} \sigma_i$ and $h^{c'}= \sum_i J_{0' i} \sigma_i$ are the cavity fields acting on $\sigma_0$ and $\sigma_{0'}$ respectively. The term $ J_{00'} \sigma_{0}\sigma_{0'}$ corresponds to the interaction between the two spins where the cavity has been made. First of all, we start by computing the part of the susceptibility containing the correlation between the spin $\sigma_0$ and $\sigma_{0'}$: $\chi_{\text{nl}} = N \overline{(\langle \sigma_0 \sigma_{0'} \rangle_ -\langle \sigma_0 \rangle \langle \sigma_{0'} \rangle )^2} $, where the label `$\text{nl}$' means non-linear. %where the factor $N$ appears because we have $N^2$ terms within the brackets and each of them is of order $1/\sqrt{N}$ and $X_{\text{nl}}$ has to be finite%. 
To compute this correlation we need to keep in mind that the terms inside the bracket are of order $1/\sqrt{N}$, and then keep all the terms of this order. Before computing the averages using the cavity method we need to derive the probability density $P(h^c, h^{c'})$. To this end, we need to compute the second order moment, i.e., the 2-point correlator 
$\langle (h^c - \langle h^c \rangle) (h^{c'} - \langle h^{c'} \rangle) \rangle = \sum_{i,j} J_{0i}J_{0j} (\langle \sigma_i\sigma_j\rangle - \langle \sigma_i \rangle \langle \sigma_j \rangle)^2$ which is  of order $1/\sqrt{N}$, but this time we keep the terms of this order because we are interested in correlations that are exactly of order $1/\sqrt{N}$.
\begin{equation}
  P_N(h^c,h^{c'}) \cong \exp\left(- \frac{(h^c-h)^2}{2(1-q)} - \frac{(h^c-h')^2}{2(1-q)} +\epsilon (h^c-h)(h^{c'}-h') \right),
\label{eq:PN}
\end{equation}
where $\epsilon(h^c-h)(h^{c'}-h')$ represents the correlation term between the two fields and $\epsilon$ is a small parameter, of order $1/\sqrt{N}$. By using \eqref{eq:PN} it is possible to derive the following marginal joint probability distribution which depends on the cavity fields and explicitly on the two cavity spins:
\begin{equation*}
P_N(h^c,h^{c'},\sigma_0,\sigma_{0'}) \propto P_N(h^c,h^{c'}) \exp\left(\beta h^c \sigma_0 + \beta h^{c'}\sigma_{0'}+\beta J_{00'} \sigma_{0}\sigma_{0'} \right).
\end{equation*}
With this marginal it is finally possible to compute the susceptibility introduced above, namely $\chi_{\text{nl}}$. The computation follows the same lines as above and we only show here the final result, which is
\begin{equation*}
  \chi_{\text{nl}} = \frac{\beta^2 A^2}{1-\beta^2A} \qquad \text{with}
  \qquad A= \int \frac{\mathrm{d}h}{\sqrt{2 \pi q}} \exp\left(-\frac{h^2}{2q}\right)(1- \tanh(\beta h)^2)^2,
\end{equation*}
and shows how the non-linear susceptibility is related to the $q$ order parameter. We observe that $\chi_{\text{nl}}$ diverges as soon as $\beta^2 A = 1$ and therefore, we can make the system eventually reach this point by increasing $\beta$ and, because of this divergence, our initial assumption for the susceptibility is wrong around this point. The assumption of a finite $\chi$ is then valid only for high temperatures or, rather, as long as $\beta^2 A < 1$. This is precisely the location of the AT line. This result is thus consistent with what we mentioned above: the cavity method shown so far is equivalent to the RS approach, because also the RS solution is only valid for high temperatures. In addition we can also give a physical meaning to the RS ansatz: it corresponds to assuming that  the  2-point correlation function is small (leading to a finite $\chi$). 

\subsubsection{Derivation of the TAP equation}
Now, let's go back to the probability measure for the cavity field in the system $\Sigma_N$:
\begin{equation*}
P_{N+1}(h^c,\sigma_0) \cong \exp\left(-\frac{(h^c-h)^2}{2(1-q)} + \beta \sigma_0 h^c\right).
\end{equation*}
With the previous measure we can compute the expectation for the cavity field in the $\Sigma_{N+1}$ system
\begin{equation}
\langle h^c\rangle_{N+1} = \sum_i J_{0 i} \langle \sigma_i \rangle_{N+1} = h + \beta(1-q) \langle \sigma_0 \rangle_{N+1},
\label{eq:hcN1}
\end{equation}
and also the expectation value of $\sigma_0$ in the same system:
\begin{equation}
\langle \sigma_0 \rangle_{N+1}= \tanh\biggl(\beta \sum_i J_{0 i} \langle \sigma_i \rangle_{N} \biggr).
\label{eq:s0_bis}
\end{equation}
Multiplying Eq.~\eqref{eq:hcN1} by $\beta$ we get $\beta h = \beta \sum_i J_{0 i} \langle \sigma_i \rangle_{N+1}- \beta^2(1-q)\langle \sigma_0 \rangle_{N+1}$, which, after applying $\tanh(\cdot)$ to both sides of the equation and making use of Eq.~\eqref{eq:s0_bis}, gives rise to the TAP equation \citep{thouless1977solution},
\begin{equation*}
\langle \sigma_0 \rangle= \tanh\biggl(\beta \sum_i J_{0 i} \langle \sigma_i \rangle - \beta^2 (1-q)\langle \sigma_0 \rangle \biggr),
\end{equation*}
where we generalised the result by omitting the label $N+1$ on the averaged terms. The first term in the argument of $\tanh(\cdot)$ is the effect of all the spins except $\sigma_0$ on $\sigma_0$, while the second term is a correction  called Onsager's reaction term. Physically speaking, the reaction term arises because the presence of $\sigma_0$, when we consider the whole system without any performed cavity, affects all the other spins, and this effect is proportional to $\langle \sigma_0 \rangle$. The TAP equation as derived above is correct as long as the connected correlation between spins is small, i.e., is of the order of $1/\sqrt{N}$, which is the only assumption made to derive the equation. From the replica point of view, the assumption of small connected correlations is equivalent to a replica symmetric ansatz and then we can conclude that the TAP equation is correct only in the high temperature regime.

\section{Cavity method for diluted graph models}
\label{sec:two}
\subsection{Replica symmetry breaking and pure states}
The cavity method applied the SK model, within replica symmetry assumptions, assumes
that the two-point correlation function between spins is small, i.e.,
$c_{ij} =\langle\sigma_{i}\sigma_{j}\rangle-\langle\sigma_{i}\rangle\langle\sigma_{j}\rangle$ is of the order of $1/\sqrt{N}$.
When the system falls into the spin glass phase, the configuration
space decomposes into many \emph{pure states}. The probability of a given configuration $\sigma = (\sigma_1,\dotsc,\sigma_N)$ can then be
decomposed as a sum over pure states,
\begin{equation*}
P(\sigma) = \sum_{\alpha}w_{\alpha}\mu_{\alpha}(\sigma),
\end{equation*}
where $\mu_{\alpha}(\cdot)$ is the measure within the pure state, which determines how configurations are weighted in one particular pure state, and $w_{\alpha}$ is the weight of the pure state $\alpha$, given by
\begin{equation*}
w_{\alpha} =\frac{\mathrm{e}^{-\beta N f_{\alpha}}}{\sum_{\alpha^{\prime}}\mathrm{e}^{-\beta Nf_{\alpha^{\prime}}}} \; .
\end{equation*}
where $f_\alpha$ called \emph{the free energy density} of the pure state $\alpha$. (Some authors prefer to use the free entropy, defined as $\phi=\log(w)/N=-\beta f$.) 
Physical quantities depend on the pure state $\alpha$ the system is in. For instance, the single spin magnetization at the pure state $\alpha$ is
\begin{equation*}
  \langle\sigma_{i}\rangle_\alpha = \sum_{\sigma \in \alpha} \sigma_i \mu_\alpha(\sigma).
\end{equation*}
More in general, the average value of any observable $\mathcal{O}$ within the pure state $\alpha$ is given by $\langle \mathcal{O} \rangle_{\alpha} = \sum_{\sigma \in\alpha} \mathcal{O}(\sigma) \mu_{\alpha}(\sigma)$.
The average magnetization over all the pure states is simply the weighted sum 
\begin{equation*}
\langle\sigma_{i}\rangle =\sum_{\alpha} w_{\alpha} \langle\sigma_{i}\rangle_{\alpha}\;.
\end{equation*}
The decomposition in pure states is justified because the escape time from a pure state grows exponentially long with the system size $N$. 

In the replica method showed in Parisi's lectures, we saw that pure states are grouped hierarchically. At the 1-step replica symmetry breaking (1RSB) level, all the states are equally seperated from each other, i.e., the overlap between two replica systems in any two different pure states is the same.  At the 2RSB level, some pure states are closer than others, forming a larger cluster structure, but the distance between any two larger clusters of pure states is the same. This hierarchical structure is also present in the cavity method. Instead, one assumes that within a pure state $\alpha$ the correlation $c_{ij}$ is weak at the 1RSB level, while the overall correlation may be strong.

If we know one pure state, we can use a set of external auxiliary fields 
$\{B^\alpha_i\}$ to quench the system into a particular pure state $\alpha$. In that case, the measure within the pure state $\alpha$ is obtained as the limit, when $ B_{i}^{(\alpha)}$ goes to $0$, of 
\begin{equation*}
P_{B_{\alpha}}(\sigma) \cong \exp \Bigl[\beta\sum
J_{ij}\sigma_{i}\sigma_{j}+\sum_{i}B_{i}^{(\alpha)}\sigma_{i}\Bigr].
\end{equation*}
The cavity method at RS level, as showed in Section~\ref{sec:one}, can be applied within a given pure state. The self-consistency equation for the magnetization is 
\begin{equation*}
  \langle\sigma_{i}\rangle_{\alpha} =
  \tanh\biggl[\beta\sum_{j}J_{ij}\langle\sigma_{j}\rangle_{\alpha}-\beta(1-q)\langle\sigma_{i}\rangle_{\alpha}\biggr].
\end{equation*}
One can write all the above equations for  each pure state and the problem will be solved at 1RSB level. However, we know nothing about the details on pure states except that they exist. Fortunately, this fact, together with the weak correlation assumption within a pure state, is enough to write a self-consistency equation of 1RSB cavity method.

Solving the SK model at 1RSB and 2RSB levels can be done, although it is rather involved~\citep{mezard1986sk}. The intricate part is that one needs to deal with the reshuffling of the pure stats weights after adding one node into the $N-$system.
\[
\left\{ w_{\alpha}^{(N)}\right\} \longrightarrow\left\{ w_{\alpha}^{(N+1)}\right\}
\]
Solving the self-consitency equation of the cavity method, finally, is the same equation got from the saddle point equation in replica method.

In this lecture, another type of system is used to illustrate the cavity method, the dilute graph model, which has a wide application on random constraint satisfaction problems. The 1RSB of such system is stable, so there is no need for a higher level symmetry breaking.

\subsection{Counting the pure states at 1RSB level}
\label{subsec:complexity}
Let's denote by $\Omega(f)$ the number of pure states with weight $w=\mathrm{e}^{-\beta N f}$. In the large $N$ limit, we are interested in its leading exponential
order, which we assume to be of the form:
\begin{equation}
\Omega(f) = \mathrm{e}^{N\Sigma(f)},\label{eq:def_complexity}
\end{equation}
where $\Sigma(f)$ is the \emph{complexity}, or \emph{configurational entropy}. 

Define the \emph{grand partition function} with a re-weighting parameter $m$
of pure states
\begin{equation*}
  \mathcal{Z}(m,\beta) = \sum_{\alpha}\exp\bigl(-\beta m N
  f_{\alpha}\bigr) = \int \exp\Bigl(N[\Sigma(f)-\beta m
  f]\Bigr)\,\mathrm{f} = \mathrm{e}^{N\Phi(m,\beta)} \;,
\end{equation*}
where $\Phi(m,\beta)$ is called \emph{grand free entropy}. As $N\rightarrow\infty$, the above integral is dominated by the largest exponential term.
\begin{align} 
    f^\ast&= \arg\max_f\left[\Sigma(f)-\beta mf\right]\notag \\
    \Phi(m,\beta)&=\Sigma(f^\ast)-\beta mf^\ast \label{eq:complexity1}
\end{align}
$\Phi(m,\beta)$ is the Legendre transform of $\Sigma(f)$. For a given $m$,
$\beta$, $\Phi(m,\beta)$ can be derived with the 1RSB cavity method.
It is assumed that $\Sigma(f)$ is a concave function. 
The complexity $\Sigma(f)$ can then be computed with an inverse Legendre transform. 
We can also compute the average free energy density over all the pure states, which is equal to the dominating value $f^\ast$. The complexity $\Sigma(f)$ can then be obtained from Eq.~\eqref{eq:complexity1}.

From a physical standpoint, we should require the complexity to be non-negative,
because otherwise there would be an exponentially small number of pure states with free energy density $f$. In the large $N$ limit, that would mean no such pure states at all. In any case, the grand partition function is dominated by the existing pure state with largest weight $w$, i.e., with the smallest free energy density. 
The phenomenon by which the measure is dominated by sub-exponentially many states is called \emph{condensation}.

The original system $Z(\beta)$ is related to $\mathcal{Z}(m,\beta)$ at $m=1$ 
if $\Sigma(f^\ast)\geq0$, where $f^\ast$ satisfies 
$$\left. \frac{\mathrm{d}\Sigma}{\mathrm{d}f}\right|_{f^\ast}=\beta m \;.$$ 
If $\Sigma(f^\ast)<0$, the original system should correspond to the largest $m$ such that $\Sigma(f^\ast)=0$.
We are left with two 1RSB phases. When $\Sigma(f^{\ast}) > 0$ we are in the so-called dynamic 1RSB (\emph{cluster phase}), and the system is dominated by exponentially many pure states. When $\Sigma(f^{\ast}) = 0$, we are in the static 1RSB (\emph{condensed phase}), and the system is dominated by sub-exponentially many pure states.

Computing the complexity is analogous to computing the entropy of a new system in which each microstate (each configuration) is a pure state $\alpha$, and where the free energy of the microstate is $f_{\alpha}$. The computation of the complexity versus the free energy density of pure states by a Legendre transform is the topic of Large deviation theory. A general review on this subject can be found in~\citep{Touchette2009}. 

\subsection{Randomly diluted graphical models}
\begin{figure}[htbp]
  \centering
  \includegraphics[scale=0.6]{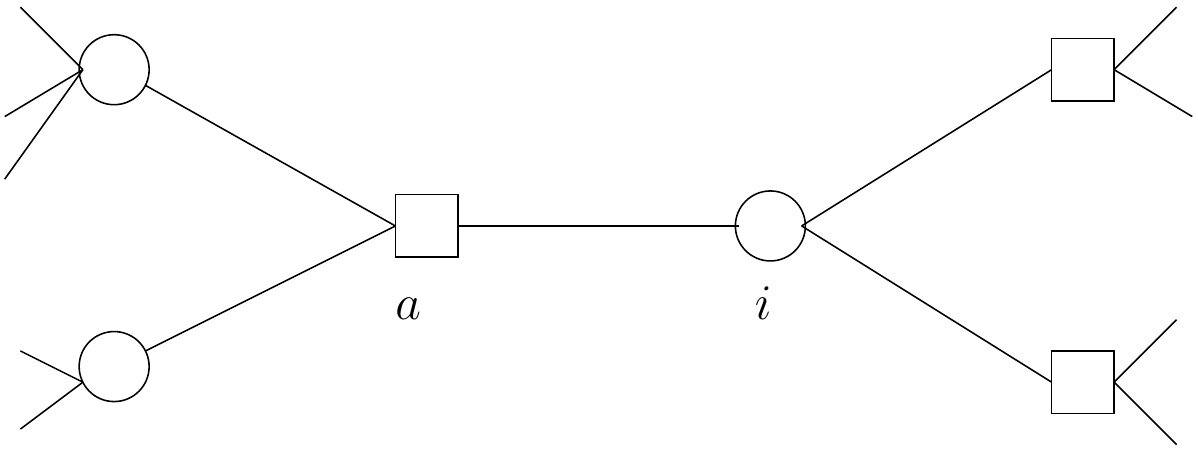}
  \caption{Factor graph: A circle node represents a variable node. A square represents
  a factor node.}
  \label{fig_factor_graph}
\end{figure}
The factor graph $\mbox{\ensuremath{\mathcal{F}}}(V,F,E)$ is a bi-partite graph  with two type nodes: variable nodes and factor nodes. Each variable node is associated
with a random variable $x_{i}$, $i\in V$, and each factor node is associated with a factor, a non-negative function $\psi_{a}(x_{\partial a})$, where $a=1,\dotsc,F$ and
$\partial a$ represents the set of neighbor variable nodes of the factor node
$a$. 

The joint probability of $x=(x_1, \dotsc, x_N)$ is expressed as
\begin{equation}
p(x)=\frac{1}{Z}\prod_{a\in F}\psi_{a}(x_{\partial a}),
\end{equation}
where $Z$ is the partition function. In such context, we may want to answer
different questions. For example, we may want to compute the marginal
probability $p_{i}(x_{i})=\sum_{x_{V\setminus\{i\}}}p(x)$. Another example
would be determining the partition function $Z$ or, rather, its first leading
exponential order, $\phi=\frac{1}{N}\log Z$. We might also want to find a
particular configuration of the variables such that $p(x)\neq0$, which is the
situation encountered in constraint satisfaction problems. 

\subsubsection*{Examples}
\begin{enumerate}[itemsep=0pt]
  \item Ising spin glass: $x_{i}\in\{+1,-1\}$, $a=(i,j)$, where $(i,j)$
    is an edge of the lattice. $\psi_{a}=\mathrm{e}^{\beta J_{ij}x_{i}x_{j}}$
  \item Coloring problem: Given a set of $q$ colors and a graph $\mathcal{G}(V,F)$,
    label each node with a color $x_{i}\in\{1,2,\dotsc,q\}$, such that no 
    neighboring nodes have the same color. Each constraint is defined on 
    the edges and has the form $\psi_{(ij)} = 1-\delta_{x_{i},x_{j}}$,
    or the soft constraint version $\psi_{(ij)}=\mathrm{e}^{-\beta\delta_{x_{i},x_{j}}}$. The inverse temperature $\beta$ alters the tolerance to the presence
    of neighbor nodes sharing the same color.
  \item $K$-SAT problem: Given $N$ boolean variables $x_{i}\in\{0,1\}$, with $i = 1,2,\dotsc,N$, and $M$ $K$-clauses in conjunctive norm form (a $K$-clause is a logical expression involving $K$ variables, or their negation, which are connected with logical \texttt{OR}s), find an assignment of boolean variables $\{x_{i}\}$ that satisfies all the $M$ clauses. In the corresponding graphical model, the factor is an indicator function, which is 1 when the clause is satisfied, and is 0 otherwise. In other words, $\psi_{a}(x_{\partial a}) = \mathbb{I}[\text{clause $a$ is satisfied]}$. We will study $K$-SAT problems in more detail in Section~\ref{sec:three}
\end{enumerate}

\subsubsection*{The structure of a factor graph}
\begin{enumerate}
\item Line or cylinder: This case can be solve exactly by the transfer matrix method. 
\item Tree: BP or cavity method is exact on tree.
\item Random hypergraph: An extension of random Erd\H{o}s-Renyi graph into
factor graph. There are $N$ variable nodes, and $M$ factor nodes. The
factor node has a fixed degree $K$, which is randomly chosen from
$N \choose K$ $K$-tuples. The degree of variable node follows the Poisson
distribution $P_{c}(d) = c^{d}\mathrm{e}^{-c}/d!$.
The length of a typical loop is of the order of $\log N$
\end{enumerate}

\subsection{Cavity method at the RS level, for general graphical models}
\subsubsection{Calculating the marginal distribution}
We consider a random hypergraph with the $N$ variables and $\alpha N$ factors, where $\alpha$ is the \emph{constraint density} in $K$-SAT. The system with $N+1$ variable nodes is generated by adding a new variable $x_{0}$ and $d$ factors, where $d$ is a random integer drawn from a Poisson distribution with mean $c=\alpha K$, the mean degree of a variable node. Each new factor is connected to $x_{0}$,
and $(K-1)$ variables randomly chosen from the $N$-variables system. Note that the constraint density $\alpha$ of $N+1$ system is slightly changed. While it does not affect the marginal distribution, it should be taken into account when computing the free energy density.

The assumption of the cavity method states that the joint probability of a
constant number of variables chosen randomly is factorized, because
the typical distance between any two variable nodes is of  order of $\log N$.

\begin{equation}
P(x_{i_{1}},x_{i_{2}},\dotsc,x_{i_{d(K-1)}}) \approx \prod_{j=1}^{d(K-1)}P(x_{i_{j}})\label{eq:bp_approx}.
\end{equation}

\begin{figure}[htbp]
  \centering
  \includegraphics[scale=0.6]{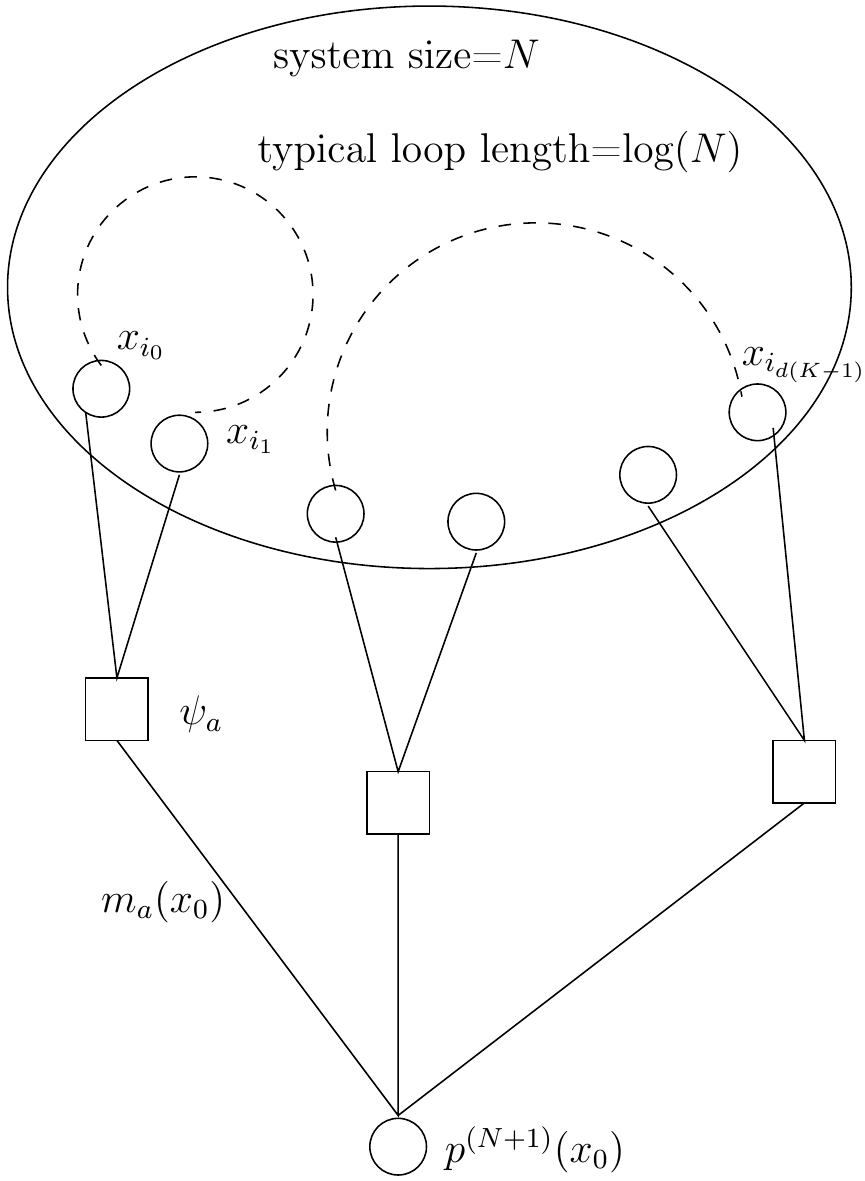}
  \caption{Illustration of the cavity method}
\end{figure}

The joint marginal probability of $x_{0}$ and the $d(K-1)$ variables
connected to the new $d$ factors is 
\begin{equation*}
  \begin{split}
    &P^{(N+1)}(x_{0},x_{i_{1}},x_{i_{2}},\dotsc,x_{i_{d(K-1)}}) \\
    &\cong \prod_{a=1}^{d}
    \psi_{a}(x_{0},x_{i_{a(K-1)+1}},x_{i_{a(K-1)+2}},\dotsc)
    P^{(N)}(x_{i_{1}}, x_{i_{2}},\dotsc,x_{i_{d(K-1)}})\\
    &\approx \prod_{a=1}^{d}\left[\psi_{a}(x_{0},x_{i_{a(K-1)+1}},x_{i_{a(K-1)+2}},\ldots)\prod_{k=a(K-1)+1}^{(a+1)(K-1)}P_{i_{k}}^{(N)}(x_{i_{k}})\right].
 \end{split}
\end{equation*}
The marginal probability $P^{(N+1)}(x_{0})$ of the newly added variable
is
\begin{equation*}
P^{(N+1)}(x_{0}) \cong \prod_{a=1}^{d}\hat{m}_{a}(x_{0}),
\end{equation*}
where
\begin{equation*}
  \begin{split}
    &\hat{m}_{a}(x_{0}) \\
    &\cong \sum_{x_{i_{a(K-1)+1}},\dotsc, x_{i_{(a+1)(K-1)}}}\psi_{a}(x_{0},x_{i_{a(K-1)+1}},\ldots,x_{i_{(a+1)(K-1)}})\prod_{k=a(K-1)+1}^{(a+1)(K-1)}P_{i_{k}}^{(N)}(x_{i_{k}}).
  \end{split}
\end{equation*}

The system with $N$ variable nodes can be considered as a system with $N+1$ variable nodes in which one node $x_{i}$ is absent. The \emph{cavity probability} $m_{i\to a}(x_i$) denotes the marginal probability of $x_{i}$, when the factor node $a$ is absent. $P_{i_{k}}^{(N)}(x_{i_{k}})$ can be considered as the cavity probability in the system with $N+1$ variables when the node $x_{0}$ and its neighboring factor nodes are absent. The self-consistent equations of the cavity probabilities are obtained by considering that the $x_{0}$ node is also a cavity node when one of its neighbor variables and neighbor factors are absent,
\begin{align}
\hat{m}_{a\to i}(x_{i}) & \cong \sum_{x_{\partial a\setminus\{i\}}}\psi_{a}(x)\prod_{j\in\partial a\setminus\{i\}}m_{j\to a}(x_{j}),\label{eq:bp1}\\
m_{i\to b}(x_{i}) & \cong \prod_{a\in\partial i\setminus\{b\}}\hat{m}_{a\to i}(x_{i}).\label{eq:bp2}
\end{align}
These equations are the same as the Belief Propagation equations, but here messages are cavity probabilities. The marginal probability of a node $x_{i}$ is then expressed as the cavity
probability 
\begin{equation*}
m_{i}(x_{i}) \cong \prod_{a\in\partial i}\hat{m}_{a\to i}(x_{i}).
\end{equation*}

\subsubsection{The Bethe free energy}
\label{sec:bethe}
The Bethe free energy can be derived by the cavity method by considering the free energy shift $f_{i+\partial i}$ when add a variable $i$ and its neighbor factors $a\in\partial i$. One has to be careful, though,  because the constraint density $\alpha$ will slightly change. This effect is eliminated by substracting $(K-1)$ times of the free energy shift $f_a$ when add a single factor $a$. For a given instance, the Bethe free energy is
\begin{equation}
  Nf=\sum_i f_{i+\partial i} - (K-1)\sum_a f_a
  \label{eq:Bethe1}
\end{equation}
One can also understand above equation in the way that the free energy shift of adding a factor $a$ is included $K$ times, when calculating the free energy shift of adding the neighbor variable $i\in\partial a$ and all $i$'s neighbor factors. So it should be substracted by $(K-1)$ extra effect.

The RS cavity independent assumption postulates that, when removing a node $i$ and 
its neighbor factor $a\in\partial i$, the partition function of the cavity system with fixed cavity variable $x_j$ $j\in \partial a \setminus i$, $a\in\partial i$ can be factorized by 
\begin{equation*}
  Z_{\setminus {i,\partial i}}(x_{j:j\in\partial a \setminus i, a\in\partial i})
  \approx
  \prod_{a\in\partial i} \prod_{j\in\partial a \setminus i} Z_{j\to a}(x_j) \;.
\end{equation*}
Here $Z_{j\to a}(x_j)$ is the partition function of the sub-system connected to $x_j$ with fixed value $x_j$ when the factor $a$ is absent.

The free energy shift $f_{i+\partial i}$ of adding a node $i$ and its neighbor factors is
\begin{equation}
  \begin{split}
  f_{i+\partial i} &=
  -\frac{1}{\beta}\log
  \frac{Z}{Z_{\setminus {i,\partial i}}}
  =
  -\frac{1}{\beta}\log
  \frac{ \sum_{x_i,x_{j:j\in\partial a \setminus i, a\in\partial i}}
           \prod_{a\in\partial i} 
           \left[
             \psi_a(x_{\partial a})
               \prod_{j\in \partial a \setminus i} Z_{j\to a} (x_j)
           \right]
        }
        {\sum_{x_{j:j\in\partial a \setminus i, a\in\partial i}}
           \prod_{a\in\partial i} 
           \prod_{j\in \partial a \setminus i} Z_{j\to a} (x_j)
        } \\
    & = -\frac{1}{\beta}\log \sum_{x_i}\left\{
      \prod_{a\in \partial i} \left[
          \sum_{x_{\partial a \setminus i} }
             \psi_a(x_{\partial a})
             \prod_{j\in \partial  a \setminus i}
             \frac{
                Z_{j\to a} (x_j)
             }{\sum_{x_j^\prime } Z_{j\to a} (x_j^\prime)}
        \right]
    \right\} \\
    & =
    -\frac{1}{\beta}\log
    \sum_{x_i}\left\{
      \prod_{a\in \partial i} \left[
          \sum_{x_{\partial a \setminus i} }
             \psi_a(x_{\partial a})
               \prod_{j\in \partial a \setminus i} m_{j\to a} (x_j)
        \right]
    \right\}.
    \label{eq:fipa}
  \end{split}
\end{equation}
Similarly, the free energy shift $f_a$ caused by adding node factor node $a$ is
\begin{equation}
  \begin{split}
    f_a & = -\frac{1}{\beta}\log \frac{Z}{Z_{\setminus a}}
    = -\frac{1}{\beta}\log \frac{
      \sum_{x_{\partial a}}  \psi_a(x_{\partial a})  
      \prod_{j\in \partial a}  Z_{j \to a} (x_j)
    }{ \sum_{x_{\partial a}}   
      \prod_{j\in \partial a}  Z_{j \to a} (x_j)} \\
    & = -\frac{1}{\beta}\log
      \sum_{x_{\partial a}}  \psi_a(x_{\partial a})  
      \prod_{j\in \partial a}  
      \frac{Z_{j \to a} (x_j)}{\sum_{x_j^\prime}
         Z_{j\to a} (x_j^\prime)}\\
    & =
      -\frac{1}{\beta}\log \sum_{x_{\partial a}}  \psi_a(x_{\partial a})  
      \prod_{j\in \partial a}  m_{j \to a} (x_j)
      \label{eq:fa}
  \end{split}
\end{equation}
Now, the Bethe free energy can be computed with Eq.~\eqref{eq:Bethe1}. 
The expression of the Bethe free energy has several variants, for example:
\begin{equation}
Nf=\sum_{i}f_{i}+\sum_{a}f_{a}-\sum_{(ia)}f_{ia},
\label{eq:Bethe2}
\end{equation}
where
\begin{align}
  f_{i} &= -\frac{1}{\beta}\log\sum_{x_{i}}\prod_{a\in\partial i}\hat{m}_{a\to i}(x_{i}),\label{eq:f1}
%  \\
%f_{a} &= -\frac{1}{\beta} \log\sum_{x_{\partial a}}\psi_{a}(x)\prod_{i\in\partial
%a}m_{i\to a}(x_{i}),\label{eq:f2}
  \\
f_{ia} &= -\frac{1}{\beta} \log\sum_{x_{i}}m_{i\to b}(x_{i})\hat{m}_{i\to b}(x_{i}).\label{eq:f3}
\end{align}
One can proof that the two Bethe free energy expressions in Eqs.~\eqref{eq:Bethe1} and \eqref{eq:Bethe2} are equivalent when the cavity probability satisfies Eqs.~\eqref{eq:bp1}--\eqref{eq:bp2}.

A comprehensive derivation of Bethe free energy by the cavity method can be found in~\citep{Mezard2003}, which also shows the 1RSB cavity method in a special simple case (the temperature $T=1/\beta$
and Parisi parameter $m$ are both 0). A review is \citep{mezard_montanari_2009}.

\subsubsection*{Average over the disorder and the graph ensemble}
To calculate the free energy average over the disorder and the graph ensemble, one should solve a self-consistent integral equations on the distribution of the cavity probabilities $P[m]$ and $\hat{P}[\hat{m}]$ 
\begin{align*}
    P[m] &= \sum_{d=1}^\infty P_c(d) \int 
    \prod_{a=1}^d  
    \left[
      \mathrm{d} { \hat{m}_a \hat{P}(\hat{m}_a)}
    \right]
    \delta
    \left[
      m- \mathbf{m}_{i\to b} [ \{ m_a\} ]
      \right], \\
      \hat{P}[\hat{m}]&=
      \int \mathrm{d} \psi_a  P_J(\psi_a)
      \int \prod_{i=1}^{K-1}
      \left[ \mathrm{d}  m_i P(m_i)
      \right] \delta
      \left[ \hat{m}- \mathbf{\hat{m}}_{a \to i} [\psi_a, \{ \hat{m}_i\} ] \right],
\end{align*}
where $\mathbf{m}_{i\to b}$ and $\mathbf{\hat{m}}_{a \to i}$ are the functionals of the BP equations~\eqref{eq:bp1} and \eqref{eq:bp2}, respectively, and $P_c(d)$ is the degree distribution of a cavity variable node, which is still a Poisson distribution with $c=\alpha K$ for a random hypergraph. The function $P_J(\psi_a)$ is the distribution of the disorder, which depends on the concrete model. For instance, in the random $K$-SAT problem, $\psi_a$ is parametrized as $J_{a}^{i}$ randomly chosen  from $\{+1,-1\}$  with equal probability. The average free energy shift when adding a factor is given by
\begin{equation*}
  \bar{f_a}=
  \int \mathrm{d} \psi_a\,  P_J(\psi_a)
      \int
      \prod_{i=1}^{K}
      \left[
        \mathrm{d}  m_i P(m_i)
      \right]
      f_i(\psi_a,\{\hat{m}_i\})
\end{equation*}
where $f_i(\psi_a,\{\hat{m}_i\})$ is defined by 
Eq.~\eqref{eq:f1}. Other average free energy shift could be written down in the similary way.
The averge free energy density over the disorder and the graph ensemble is 
\begin{equation}
\bar{f} = \bar{f}_{i+\partial i}-\alpha(K-1)\bar{f}_{a}
\label{eq:avg_f}
\end{equation}

In general it is hard or impossible to get an analytical solution of above equation, but one can
use numerical simulations to solve it. The algorithm is called
\emph{Population Dynamics}, or density evolution.\label{popdyn}
% \begin{algorithmic}[H]
  % \STATE Initialize $\{m_{i\to a}\}$
  % \FOR{$t=1,\dotsc,T$}
  % \FOR{$i=1,\dotsc,N$}
  % \STATE Draw an integer $d$ according to degree Poisson distribution.
  % \ENDFOR Draw $d-1$ integers uniformly from $\{1,\dotsc,N\}$
  % \ENDFOR
% \end{algorithmic}

\vspace*{-3mm}
\begin{center}
  \begin{minipage}[t]{0.85\textwidth}
\small
    \vspace{0pt}
    \emph{Initialization:} Set an array $P$ to store the messages $\{m_{i\to a}\}$.
(Note that if $x_{i}$ is Ising variable, $m_{i\to a}(x_{i})$ can
be parametrized by a single real number).
\begin{enumerate}[itemsep=0pt,leftmargin=6mm]
\item An integer $d$ is randomly assigned following the Poisson distribution $d\sim P_{c}(d)$
\item Pick $(K-1)d$ messages randomly from the array $P$
\item Generate $d$ $\psi_{a}$'s following $P_J(\psi_a)$.
\item Compute a new message $\hat{m}^{\prime}$ with Eqs.~\eqref{eq:bp1}--\eqref{eq:bp2},
and compute $f_{i+\partial i }$ with Eq.~\eqref{eq:fipa}.
\item Choose a message randomly in $P$ and replace it by the new one $\hat{m}^{\prime}$
\item Pick K messages randomly from the array $P$, and generate a factor $\psi_a$ following $P_J(\psi_a)$. Compute $f_a$ with Eq.~\eqref{eq:fa}.
\item Repeat 1--5 until getting a stable distribution $P(\hat{m})$. Then, keep
repeating 1--6 to get the mean $\bar{f}_{i+\partial a},\bar{f}_{a}$,
and calculate $\bar{f}$ with Eq.~\eqref{eq:avg_f}
\end{enumerate}
  \end{minipage}
\end{center}
For more discussion on BP free energy on average cases, one can refer
to~\citep{mezard_montanari_2009}, pages 322--325.

\subsection{Cavity method at 1RSB level}
Something may go wrong for the Bethe independent hypothesis Eq.~(\ref{eq:bp_approx}),
and there are two potential reasons for this. The first possibility is that Eq.~\eqref{eq:bp_approx} holds only when the size of system is infinitely large, $\log(N) \rightarrow \infty$. For a finite  system Eq.~\eqref{eq:bp_approx} is only an approximation. The other possible reason is that, when the constraint density $\alpha$ is high or the temperature is low, the Bethe hypothesis may fail even for an infinitely large system.
%, i.e. $P(x_{1},x_{2},\dotsc,x_{n}) \neq p_{1}(x_{1}) p_{2}(x_{2}) \dotsb p_{n}(x_{n}).$ 
For this latter case the whole probability distribution does not longer factorize,
$P(x_{1},x_{2},\dotsc,x_{n}) \neq p_{1}(x_{1}) p_{2}(x_{2}) \dotsb
p_{n}(x_{n})$, and so we need to make a more accurate assumption. As proposed in \citep{Mezard2001}, we invoke the 1RSB approximation, by which the probability distribution factorizes within each pure state $\alpha$, but not globally. More specifically, because of the presence of pure states, the whole Gibbs measure splits into many states $\alpha$, and within the measure $\mu_{\alpha}(\cdot)$
of a pure state, the independent hypothesis still holds:
\begin{equation}
\mu_{\alpha}(x_{1},x_{2},\ldots,x_{n})\approx\mu_{\alpha}(x_{1})\mu_{\alpha}(x_{2})\ldots\mu_{\alpha}(x_{n}).
\label{eq:1rsb_assumption}
\end{equation}

\begin{figure}[htbp]
  \centering
  \includegraphics[scale=0.6]{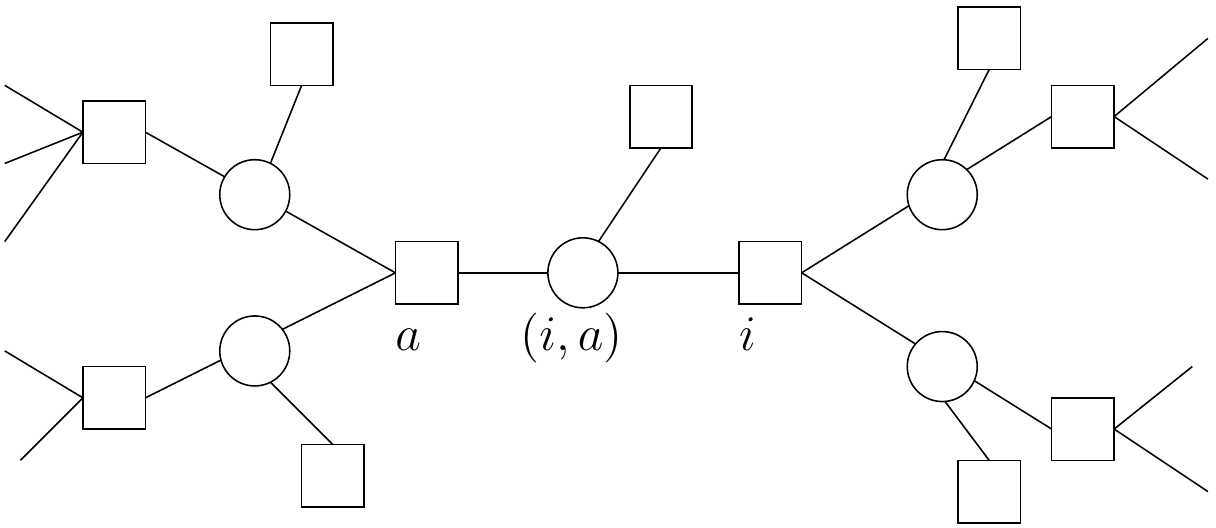}
  \caption{Computing the grand partition function by a new graphical model.}
  \label{1rsb_graphical_model}
\end{figure}

Furthermore, it is assumed that the number of pure states and fixed points of BP solutions are the same up to the first exponential leading order. The leading exponential order of the number of pure states with free energy density $f$ is $\Sigma(f)$, as defined in Eq.~\eqref{eq:def_complexity}. The grand partition function is expressed as:
\begin{align*}
\mathcal{Z}(m,\beta) & =\sum_{\alpha}\mathrm{e}^{-\beta mNf_{\alpha}}\\
 & =\sum_{\{m_{i\to a}\text{'s are fixed point}\}}\mathrm{e}^{-\beta mNf_{\alpha}[\{m_{i\to a}\}]}\\
 & =\int_{\{m_{i\to a},\hat{m}_{a\to i}\}}d\{m_{i\to a}\}d\{\hat{m}_{a\to i}\}\prod_{(i,a)}\delta\left[m_{i\to a}-\mathbf{m_{i\to a}}[\{\hat{m}_{\text{input msgs}}\}\right]\nonumber \\
 & \qquad\prod_{(i,a)}\delta\left[\mathbf{\hat{m}_{a\to i}}-\hat{p}_{a\to i}[m_{\text{input msgs}}]\right]\prod_{i}\mathrm{e}^{-\beta mf_{i}[\cdot]}\prod_{a}\mathrm{e}^{-\beta mf_{a}[\cdot]}\prod_{(i,a)}\mathrm{e}^{\beta mf_{(ia)}[\cdot]}
\end{align*}
where $\mathbf{\hat{m}_{i\to a}}[\cdot]$ , $\mathbf{m_{a\to i}}[\cdot]$ are the functionals defined by Eqs.~\eqref{eq:bp1}--\eqref{eq:bp2}, and $f_{i}[\cdot]$, $f_{a}[\cdot]$, and $f_{(ia)}[\cdot]$ are defined in Eqs.~\eqref{eq:f1}--\eqref{eq:f3}. The delta function ensures that the messages satisfy the BP iteration, Eqs.~\eqref{eq:bp1}--\eqref{eq:bp2}, so the integral means that it sum over all the BP fixed point with the weight $w=\mathrm{e}^{-\beta mf_{\text{BP}}}$.

Above expression is precisely an another graphical model defined on a new factor graph, showed in Fig.~\ref{1rsb_graphical_model}. The joint probability is still factorized and defined on the factor graph with the same topological structure. So the sparsity condition of the graph still holds. The Bethe approximation on the new graphical model is the assumption of 1RSB cavity method. Computing the graph partition function, the complexity, or any other physical quantity, goes along the same lines as the cmputations at RS level. The only difference is that now the variables we operate with are functions (a cavity probability at RS level), and factors are functionals. More details on 1RSB cavity method can be found in Chapter 19 of \citep{mezard_montanari_2009}.

\section{An example: Random K-SAT problem}
\label{sec:three}
\subsection{Cavity Method and Random K-satisfiability}
In the previous section we saw that replica symmetric (RS) cavity method leads to Belief Propagation (BP) equations, and that we can average the BP equations to get the density evolution description of the BP equation. We also saw that, at an abstract level, the 1RSB is associated with the proliferation of states, and that there is a whole hierarchy of such transitions. 

In this section we will show how the cavity method works in practice. Although the cavity method has been used in the Sherrington-Kirkpatrick model up to two-step replica symmetric breaking (2RSB)~\citep{mezard1986sk}, the derivation becomes too technical and is not particularly enlightening. The random $K$-SAT problem provides another, more workable example in which to use of message-passing techniques. We'll start with a short summary of the problem, to set the notation.

\subsubsection{Definitions and notation}
We consider $N$ boolean variables $x_i \in \{0, 1\}$, with $i=1,\ldots, N$. In our representation the value $0$ corresponds to `false', while the value `1' corresponds to `true'. A satisfiability problem is defined as a set of logical constraints that these random variables have to satisfy. Each logical constraint is called a \emph{clause}, and is expressed as a logical \texttt{OR} of a subset of the boolean variables that may or not be negated. The negation of variable $x_i$ is denoted by $\bar{x}_i \equiv 1 - x_i$.  An example of 2-clause is ``either $x_1$ is true or $x_2$ is false'', expressed more succintly as $x_1 \vee \bar{x}_2$, where $\vee$ denotes the logical \texttt{OR}. Another example is the 3-clause $x_1 \vee x_2 \vee \bar{x}_3$, which is satisfied by all configurations of $x_1, x_2, x_3$ except for $\{x_1 =0, x_2 = 0, x_3 =1\}$. In general, a satisfiability problem consists of a set of $M$ clauses $C_1, C_2, \ldots, C_M$ that have to be satisfied simultaneously. The problem is satisfiable if there is at least one choice of the boolean variables $x=(x_1, \ldots, x_N)$, also called an \emph{assignment},  that satisfies the logical formula
\begin{equation}
  F = C_1 \wedge C_2 \wedge \cdots \wedge C_M,
  \label{eq:F}
\end{equation}
where $\wedge$ is the logical \texttt{AND}.

In a $K$-SAT problem, each clause consists of exactly $K$ variables. We consider \emph{random} $K$-SAT problems, where each clause $C_a$, $a=1,\ldots, M$, contains exactly three variables chosen randomly in $\{x_1,\dotsc,x_N\}$, and each variable is negated randomly with probability $1/2$. In other words, each clause is drawn with uniform distribution from the set of all the ${N \choose K} 2^{K}$ clauses of length $K$.

\begin{figure}[htbp]
  \centering
  \includegraphics{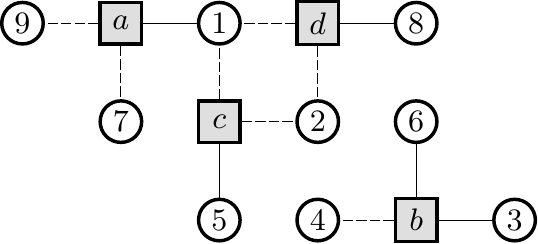}
  \caption{Example of factor graph with nine variable nodes, $i=1,\dotsc,9$ and 4 factor nodes $a,b,c,d$, The factor graph encodes the formula $F=(x_1 \vee \bar{x}_7 \vee \bar{x}_9) \wedge (x_3 \vee \bar{x}_4 \vee x_6) \wedge (\bar{x}_1 \vee \bar{x}_2 \vee x_5) \wedge (\bar{x}_1 \vee \bar{x}_2 \vee x_8)$.}
  \label{fig:3sat_graph}
\end{figure}

An instance of a $K$-SAT problem can be represented by a factor graph,  where variable nodes correspond to the boolean variables and factor nodes correspond to clauses. When the variable $x_i$ (or its negation) appears in clause $a=1,\ldots,M$, the node $i$ is connected to the clause factor $a$. It is useful to use a slightly modified version of the standard factor graph, in which the edge between $i$ and $a$ is is plotted with either a solid or a dashed line depending on whether the variable $i$ appears unnegated or negated in clause $a$ (see Fig.~\ref{fig:3sat_graph} for an example). With this modification there is a one-to-one correspondence between a $K$-SAT problem and a factor graph. For consistency, we carry over the notation and use the indices $i,j,\ldots$ for variable nodes and indices $a, b,\ldots$ for factor nodes. 

Notice that each factor node has a fixed degree $K$, but the degree of a variable node is random. More specifically, because a randomly chosen $K$-uple contains the variable $i$ with probability $K/N$, the degree of the variable node $i$ is a binomial random variable with parameters $M$ and $p=K/N$. In the limit of large $N$, the binomial distribution can be safely approximated by a Poisson disitribution with parameter $\alpha K$, i.e., $\Pr(\text{degree}_i = n) = \mathrm{e}^{-K\alpha} (K\alpha)^n/n!$. 

\begin{figure}[htbp]
  \centering
  \includegraphics{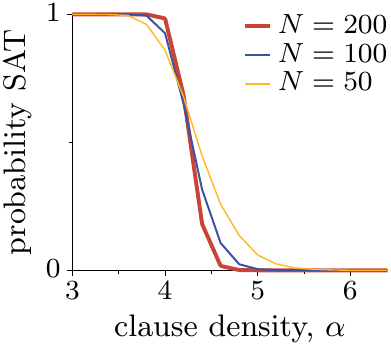} \quad\includegraphics{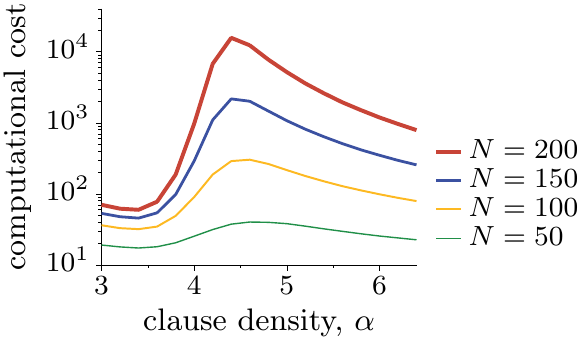}
  \caption{Left: probability that a formula drawn from the random 3-SAT ensemble is satisfiable, as a function of the clause density $\alpha=M/N$. Right: Computational time (in arbitrary units) required to either find a solution or prove that there is none, as a function of the clause density. Figures adapted from~\citep{mezard_mora_jphysiolp2009}.}
  \label{fig:3SAT}
\end{figure}

The crucial parameter that characterizes random $K$-SAT problems is the clause density $\alpha \equiv M / N$, which sets the ratio of constraints per variable. Intuitively, one expects that for small $\alpha$ most of the instances will be satisfiable, while for large enough $\alpha$ most of the instances will be unsatisfiable. Numerical experiments confirm this intuition (see Fig.~\ref{fig:3SAT}, left). The probability that a random instance is SAT drops from values close to 1 to values close to 0 as crosses the value $\alpha_c \approx 4.3$, and this transition becomes sharper the larger the number of variables $N$ is. This is the characteristic behavior of a phase transition, and as such it has been analyzed using the methods of statistical physics (some refs here).

The clause density also determines how hard the problem is. The difficulty of the problem can be quantified by the time taken by an algorithm to decide whether a typical instance is satisfiable or not. It turns out that a problem is easy when $\alpha$ is well below the critical value $\alpha_c$, it becomes harder as $\alpha$ approaches $\alpha_c$ (see Fig.~\ref{fig:3SAT}, right), and less hard when $\alpha$ is much larger than $\alpha_c$. In other words, the region around the phase transition is the hardest from a computational point of view. In the following we will define the thermodynamic limit as $M\rightarrow \infty$ and $N \rightarrow \infty$ while keeping the clause density $\alpha$ constant.

\subsection*{Belief Propagation}
Each variable $i$ appears in a random set of clauses. We denote by $\partial i$ the set of indices of the clauses where $i$ appears. In the factor graph, $\partial i$ is the set of factor nodes adjacent to the variable node $i$. Similarly, we denote by $\partial a$ the indices of the $K$ variables appearing in clause $a$, and by $x_{\partial a}$ the corresponding variables, i.e., $x_{\partial a} \equiv \{x_i \mid i\in \partial a\}$. For later convenience we define the number
\begin{equation*}
  J_{ai} = 
  \begin{cases}
    0 & \text{if}\ x_i \in C_a,\\
    1 & \text{if}\ \bar{x}_i \in C_a.
  \end{cases}
\end{equation*}
We will also distinguish the neighbors of $i$, $a \in \partial i$, according to the values of $J_{ai}$, and define $\partial_0 i = \{a \in \partial i \mid J_{ai} = 0\}$ and $\partial_1 i = \{a \in \partial i \mid J_{ai} = 1\}$.

\begin{figure}[tbp]
  \centering
  \includegraphics{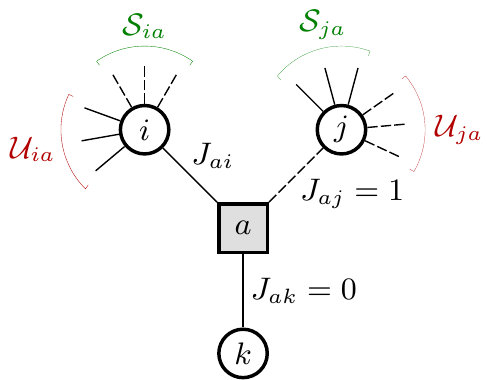}
  \caption{Factor graph associated with the single 3-clause $x_i \vee \bar{x}_j \vee x_k$. For clarity we show only the subsets $\mathcal{U}$ and $\mathcal{S}$ associated with nodes $i$ and $j$.}
  \label{fig:fg}
\end{figure}

Given the edge between the factor node $a$ and the variable node $i$, it is useful to distinguish the set of all remaining edges of $i$ according to whether or not their associated $J$s coincide with $J_{ai}$:
\begin{align*}
  \mathcal{S}_{ia} \equiv& \{b \in \partial i \backslash a \mid J_{bi} = J_{ai}\},\\
  \mathcal{U}_{ia} \equiv& \{b \in \partial i \backslash a \mid J_{bi} = 1-J_{ai}\},
\end{align*}
where $\partial i \backslash a$ means the set of all factors connected to $i$, excluding $a$. It follows from these definitions that the neighborhood of $i$ is partitioned as $\partial i = \{a\} \cup \mathcal{S}_{ai} \cup \mathcal{U}_{ai}$. Figure~\ref{fig:fg} summarizes our notation and conventions.

Given the satisfiability formula $F$ in Eq.\eqref{eq:F}, we consider the uniform probability distribution $P_{\text{\textsc{sat}}}(x)$ over the truth assingments $x=(x_1, \ldots, x_N) \in \{0,1\}^N$ that satisfy $F$, assuming they exist. This probability can be written as
\begin{equation}
    P_{\text{\textsc{sat}}}(x) \cong \prod_{a=1}^{M} \psi_a(x_{\partial a}),
    \label{eq:prob_SAT}
\end{equation}
Each factor $\psi_a(x_{\partial a})$ is 1 if clause $a$ is satisfied by the assignment $x$, and is 0 otherwise. Put differently,
\begin{equation}
  \psi_a(x_{\partial a}) = \mathbb{I}(x_{\partial a}\ \text{satisfies}\ C_a),
  \label{eq:psi_indicator}
\end{equation}
with $\mathbb{I}$ being the indicator function.

\subsubsection{The Belief Propagation equations}
Belief propagation (BP) is an iterative algorithm that operates on `messages' associated with the directed edges of a factor graph. For each edge $(i, a)$ there exist two messages $\hat{m}_{a\rightarrow i}(x_i)$, $m_{i\rightarrow a}(x_i)$, defined in the space of probability distributions on the set $\{0, 1\}$: their values lie the interval $[0, 1]$ and satisfy $\sum_{x_i} m_{i\rightarrow a}(x_i) = 1$. Messages are updated according to
\begin{align}
  \hat{m}^{(t)}_{a \rightarrow i}(x_i) & \cong   
  \!\sum_{x_{\partial a \backslash i}} \psi_a(x_{\partial a}) \prod_{k \in \partial a\backslash i}\!m^{(t)}_{k\rightarrow a}(x_k),\label{eq:BP1}\\
  m^{(t + 1)}_{i \rightarrow a}(x_i) & \cong \hspace{-1.5mm}\prod_{b \in \partial i\backslash a} \!
  \hat{m}^{(t)}_{b\rightarrow i}(x_i).\label{eq:BP2}
\end{align}
These are the belief propagation, or \emph{sum-product}, update rules. In tree-like graphical models the messages converge to fixed-point values. The resulting message $m^{(\infty)}_{i\rightarrow a}(x_i)$ is the marginal distribution of variable $x_i$ in a modified graphical model that does not include the factor $a$. Analogously, $\hat{m}^{(\infty)}_{a\rightarrow i}(x_i)$ is the marginal distribution of $x_i$ in a graphical model where all factors $\partial i$ but
$a$ have been removed. 

We can simplify the formulation of the BP equations for $K$-SAT, using the fact that variables $x_i$ are all binary to parametrize the messages with a single real number. We define
\begin{align*}
  \zeta_{ia} &\equiv m_{i\rightarrow a}(x_i = J_{ai}) \in [0, 1],\\
  \hat{\zeta}_{ai} &\equiv \hat{m}_{a\rightarrow i}(x_i = J_{ai}) \in [0, 1].
\end{align*}
From the normalization of the messages, it follows that $m_{i\rightarrow a}(x_i = 1 - J_{ai}) = 1 - \zeta_{ia}$ and $\hat{m}_{a\rightarrow i}(x_i = 1 - J_{ai}) = 1 - \hat{\zeta}_{ai}$. The variables $\zeta_{ai}$ and $\hat{\zeta}_{ia}$ can be interpreted as the message associated the wrong direction of $x_i$. In terms of $\zeta_{ai}$ and $\hat{\zeta}_{ia}$, the BP equations~\eqref{eq:BP1}--\eqref{eq:BP2} read
\begin{align}
  \hat{\zeta}_{ai} &= \dfrac{1 - \prod_{j \in \partial a \backslash i} \zeta_{ja}}{1 + \bigl(1 - \prod_{j \in \partial a \backslash i} \zeta_{ja}\bigr)},\label{eq:BPb}\\
  \zeta_{ia} &= \dfrac{\bigl[\prod_{b\in \mathcal{S}_{ia}} \!\hat{\zeta}_{bi}\bigr] \bigl[\prod_{b \in \mathcal{U}_{ia}}\! (1 - \hat{\zeta}_{bi})\bigr]}{\bigl[\prod_{b\in \mathcal{S}_{ia}} \!\hat{\zeta}_{bi}\bigr] \bigl[\prod_{b \in \mathcal{U}_{ia}}\! (1 - \hat{\zeta}_{bi})\bigr] + \bigl[\prod_{b\in \mathcal{S}_{ia}}\! (1 - \hat{\zeta}_{bi})\bigr] \bigl[\prod_{b \in \mathcal{U}_{ia}}\! \hat{\zeta}_{bi}\bigr]},\label{eq:BPb2}
\end{align}
where we use the convention that a product of zero factors is 1. The number of operations required to evaluate the right hand side of these two equations is of the order of $O(|\partial a|)$ and $O(|\partial i|)$, respectively, where $|A|$ is the cardinality of $A$. To solve Eqs.~\eqref{eq:BPb} we update the messages until a fixed point is reached, after which we can obtain the marginals.

\subsubsection{Statistical Analysis}
We can go further and use the equations to derive the overall distribution of the messages. The idea is to draw a random edge $(i, a)$ in the factor graph and consider the corresponding fixed point of the messages $\zeta_{ia}, \hat{\zeta}_{ai}$ as random variables. Within the replica-symmetric (RS) assumption, and when $N\rightarrow \infty$, these variables converge in distribution to edge-independent random variables $\zeta, \hat{\zeta}$, with distribution
\begin{align}
  \hat{\zeta} & \stackrel{d}{=} \frac{1 - \zeta_1 \dotsb \zeta_{K-1}}{2 - \zeta_1 \dotsb \zeta_{K-1}},\label{eq:zetaP} \\
  \zeta & \stackrel{d}{=} \frac{\hat{\zeta_1} \dotsb \hat{\zeta}_{p}(1- \hat{\zeta}_{p+1})\dotsb (1 - \hat{\zeta}_{p+q})}{\hat{\zeta_1} \dotsb \hat{\zeta}_{p}(1- \hat{\zeta}_{p+1}) \dotsb (1 - \hat{\zeta}_{p+q}) + (1 - \hat{\zeta_1}) \dotsb (1 - \hat{\zeta}_{p})\hat{\zeta}_{p+1}\dotsb \hat{\zeta}_{p+q}}\label{eq:hzetaP}
\end{align}
where $\stackrel{d}{=}$ means `equal in distribution'. The numbers $p$ and $q$ are two i.i.d.\ Poisson random variables with mean $K\alpha / 2$, and correspond to the random number of unnegated and negated edges in a variable node---namely, the numbers $|\partial_0 i|$ and $|\partial_1 i|$. The variables $\zeta_1,\dotsc,\zeta_{K-1}$ are i.i.d.\ copies of $\zeta$, and $\hat{\zeta}_1,\dotsc,\hat{\zeta}_{p+q}$ are i.i.d.\ copies of $\hat{\zeta}$. The probability density functions for $\zeta$ and $\hat{\zeta}$ defined by Equations~\eqref{eq:zetaP}--\eqref{eq:hzetaP} are to be understood as
\begin{align}
  p(\hat{\zeta}) & = \int \prod_{i=1}^{K-1} \bigl\{ \mathrm{d} \zeta_i \, p(\zeta_i)\bigr\} \, \delta\left(\frac{1 - \zeta_1 \dotsb \zeta_{K-1}}{2 - \zeta_1 \dotsb \zeta_{K-1}}\right),\\
  p(\zeta) & = \sum_{r=0}^{\infty} \sum_{s=0}^{\infty} P(r) P(s)\int
  \prod_{i=1}^{K-1} \bigl\{ \mathrm{d} \hat{\zeta}_i\, p(\hat{\zeta}_i)\bigr\} \notag\\ & \;\times \delta\left(\dfrac{\displaystyle\prod_{a=1}^{r} \hat{\zeta}_a \prod_{b=r+1}^{r+s} (1 - \hat{\zeta}_b)}{\displaystyle \prod_{a=1}^{r} \hat{\zeta}_a \prod_{b=r+1}^{r+s} (1 - \hat{\zeta}_b) + \prod_{a=1}^{r} (1 - \hat{\zeta}_a) \prod_{b=r+1}^{r+s}  \hat{\zeta}_b}\right), \end{align}
  where $P(r)$ is the probability distribution of a Poisson random variable $X$, $\Pr(X=r) = \mathrm{e}^{-\lambda} \lambda^{r} / r!$, with mean $\lambda = K\alpha / 2$.

  The generic way to solve the set of coupled equations~\eqref{eq:zetaP}--\eqref{eq:hzetaP} is by using \emph{population dynamics} (see p.~\pageref{popdyn}). In this numerical method one approximates the distribution of $\zeta$ (or $\hat{\zeta}$) through a sample of $N$ i.i.d.\ copies of the variable and exploits the property that, in the limit of large $N$, the empirical distribution of the sample converges to the actual distribution.

\subsection{Free Entropy}
Recall from Section~\ref{sec:two} that the free entropy informs us about the number of solutions, and it is a function of the messages of the factor graph. We now evaluate the free entropy for a $K$-SAT problem. If $E$ denotes the set of edges in the graph, there are $2|E|$ messages, which we collectively denote by $m \equiv \{m_{i\rightarrow a}(\cdot), \hat{m}_{a\rightarrow i}(\cdot)\}$. The free entropy then reads
\[F(m) = \sum_{a\in F} F_a(m) + \sum_{i\in V} F_i(m)  - \sum_{(ai) \in E}  F_{ai}(m),\]
where $F$ is the set of factor nodes, $V$ is the set of variable nodes, and 
\begin{align}
  F_a(m) & = \log\left[\sum_{x_{\partial a}} \psi_a(x_{\partial a})\,\prod_{i \in \partial a} m_{i \rightarrow a}(x_i)\right],\label{eq:Fm1}\\
  F_i(m) & =  \log\left[\sum_{x_i} \prod_{b \in \partial i} \hat{m}_{b \rightarrow i} (x_i)\right],\label{eq:Fm2}\\
  F_{ai}(m) & =  \log\left[\sum_{x_i} m_{i\rightarrow a}(x_i) \,\hat{m}_{a \rightarrow i} (x_i)\right].\label{eq:Fm3}
\end{align}
In $F_a(m)$, the sum $\sum_{x_{\partial a}}$ is over all the possible configurations of the variable nodes adjacent to $a$. In terms of $\zeta \equiv \{\zeta_{ia}, \hat{\zeta}_{ai}\}$, Eqs.~\eqref{eq:Fm1}--\eqref{eq:Fm3} read
\begin{align}
  F_a(\zeta) & = \log\Bigl[1 - \prod_{i\in\partial a} \zeta_{ia}\Bigr]\!,\label{eq:Fz1}\\
  F_i(\zeta) & =  \log\left[\prod_{a \in \partial_0 i}\! \hat{\zeta}_{ai} \prod_{b \in \partial_1 i} (1 - \hat{\zeta}_{bi}) + \prod_{a \in \partial_0 i} (1 - \hat{\zeta}_{ai}) \prod_{b \in \partial_1 i} \hat{\zeta}_{bi}\right], \label{eq:Fz2}\\
  F_{ai}(\zeta) & =  \log\Bigl[ \zeta_{ia} \hat{\zeta}_{ai} + (1 - \zeta_{ia})(1 - \hat{\zeta}_{ai})\Bigr].\label{eq:Fz3}
\end{align}

\begin{figure}[htbp]
\centering
\includegraphics{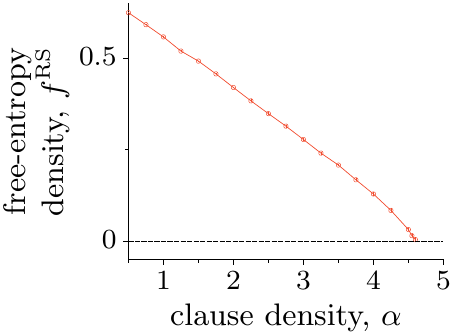}
\caption{Estimate of the Bethe free-entropy density as a function of the clause density, for 3-SAT and assuming replica symmetry. The curve reaches 0 entropy at around $\alpha_{*}(3) \approx 4.677$.}
\label{fig:entropy}
\end{figure}

In section~\ref{sec:bethe} we saw that under RS assumptions, the Bethe free-entropy density in the thermodynamic limit is
\begin{equation}
  \lim_{N\rightarrow \infty} \frac{F}{N} = f^{\text{RS}} = f^{\text{RS}}_{v}  + \alpha f_{c}^{\text{RS}}  - K \alpha f_{e}^{\text{RS}}
\label{eq:entropy}
\end{equation}
where
\begin{align*}
  f^{\text{RS}}_{v} &= \ee \log\left[ \prod_{a=1}^{p} \! \hat{\zeta}_{a} \prod_{b=p+1}^{p+q}  (1 - \hat{\zeta}_{b}) + \prod_{a=1}^{p} (1 - \hat{\zeta}_{a}) \prod_{b=p+1}^{p+q} \hat{\zeta}_{b}\right],\\
  f^{\text{RS}}_{c} &= \ee \log\left[ 1 - \zeta_1 \dotsb \zeta_{K-1}\right],\\
  f^{\text{RS}}_{e} &= \ee \log\left[ (1 - \zeta_1)(1  - \hat{\zeta}_1) + \zeta_1   \hat{\zeta}_1 \right].
\end{align*}
Here $\mathbb{E}$ denotes expectation with respect to the variables $\zeta_1, \dotsc, \zeta_K$ (the i.i.d.\ copies of $\zeta$), $\hat{\zeta}_1,\dotsc,\hat{\zeta}_{p+q}$ (the i.i.d.\ copies of $\hat{\zeta}$), and the Poisson random variables $p$ and $q$. We can use population dynamics to estimate the distributions of $\zeta$ and $\hat{\zeta}$, and then use the resulting samples to estimate the
free-entropy density, Eq.~\eqref{eq:entropy}. The outcome of this procedure, repeated for several values of $\alpha$,  is summarized in Fig.~\ref{fig:entropy}. The entropy density is strictly positive and decreasing for $\alpha \leq \alpha_{*}(K)$, with $\alpha_{*}(3) \approx 4.677$. The value $\alpha_{*}(K)$ is the RS prediction for the SAT-UNSAT threshold $\alpha_{\text{s}}(K)$, where $K$-SAT instances cease to be satisfiable.

Unfortunately, this result is inconsistent with the upper bound $\alpha_{\text{UB}}(3) \approx 4.666$, derived rigurously from the first moment method (see lecture 3 by Cris Moore). The reason for this contradiction is that the RS assumption is expected to be correct only up to the condensation transition $\alpha_c(3)\approx 3.86$, where pure states start to proliferate (see Sec.~\ref{sec:two}). 

\subsection*{BP-guided decimation}
\begin{figure}[htbp]
  \centering
  \includegraphics{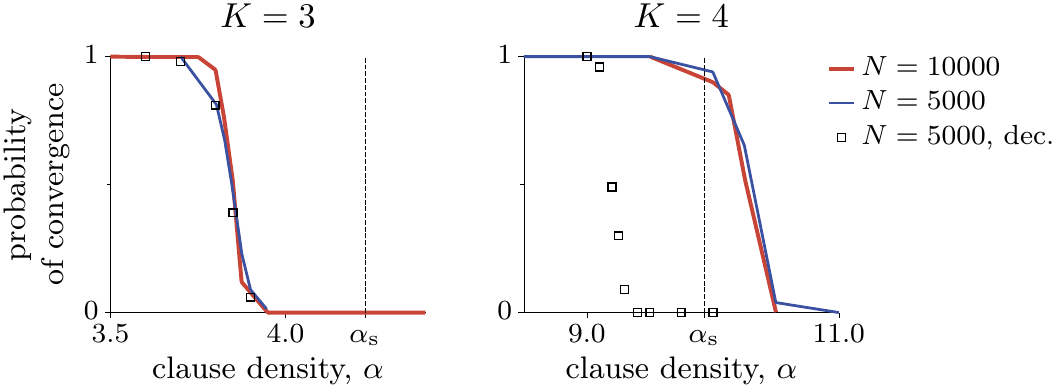}
  \caption{Empirical probability that the BP algorithm converges to a fixed point, as a function of the the clause density, for 3-SAT (left) and 4-SAT (right). The estimate is based on 100 instances with the number of variables indicated in the legend. Squares indicate emprirical probability that BP-guided decimation finds a SAT assignment, using 100 instances with 5000 variables each. The vertical dashed line is located at the  SAT-UNSAT threshold $\alpha_{\text{s}}$. Parameters of the decimation: $\delta=10^{-2}$, $t_{\text{max}} = 10^3$. Figures adapted from~\citep{mezard_montanari_2009}.}
  \label{fig:bp_conv}
\end{figure}

Another way to realize that the RS assumption cannot be valid close to the SAT-UNSAT threshold is by using the BP iteration. We can just pick a random $K$-SAT instance, initialize the messages with uniform random numbers, and then iterate the BP equations~\eqref{eq:BPb}--\eqref{eq:BPb2} until no message changes by more than some prescribed small number $\delta$. If we fix a large time $t_{\text{max}}$, we can estimate the probability of convergence within $t_{\text{max}}$ by repeating the same experiment many times. Figure~\ref{fig:bp_conv} summarizes such an experiment for $K=3$ and $K=4$. The estimated probability curves show a sharp decrease around a critical value of $\alpha$, which we denote $\alpha_{\text{\textsc{bp}}}$ and which turns out to be robust to variations of $\delta$ and $t_{\text{max}}$

We can go further and find a SAT assignment based on the messages obtained after convergence of the BP iteration. The method is called BP-guided decimation and is as follows. Given the BP estimate of the marginal of $x_i$, we compute the \emph{bias} $\pi_i \equiv P_i(1) - P_i(1)$ for each variable, and then pick the variable with highest $|\pi_i|$. This variable is fixed to its favored value (i.e., $x_i$ is set to $0$ if $\pi_i >0$, or to $1$ otherwise), and the SAT formula is reduced (decimated) using this individual assignment. The method is repeated until all the variables are assigned, or until the BP fails to converge. The probability that BP-guided decimation results in a SAT assignment is shown in Figure~\ref{fig:bp_conv}, for several values of $\alpha$ and for $K=3,4$. Note that for 3-SAT the decimation method returns a SAT assignemt almost everytime the BP iteration converges (that is, for $\alpha \lesssim 3.85$). In contrast, for 4-SAT BP-guided decimation finds SAT assignments for $\alpha \lesssim 9.25$, while BP converges most of the time for $\alpha \lesssim 10.3$ (a value that is larger than the conjectured SAT-UNSAT threshold, $\alpha_{\text{s}}(4) \approx 10.93$).

This numerical experiment shows that something goes wrong when $\alpha$ is large enough. It also shows that 4-SAT is qualitatively different from 3-SATs; what makes BP fail at large $\alpha$ differs depending on the $K$ we consider. For $K=3$ the BP fixed point becomes unstable at around $\alpha_{\text{st}} \approx 3.86$, which leads to errors in decimations. [short sketch on how to determine stability: entropic factor vs correlation decay] For $K=4$, in contrast, the BP fixed point remains stable but does not lead to the correct marginals because the 1RSB condensation threshold $\alpha_c$ is crossed. 

\subsection*{The 1RSB cavity method}
We could proceed with the strategy outlined in Section~\label{subsec:comkplexity}, using the BP approximation in the auxiliary model in order to estimate the complexity function $\Sigma(f)$. 
This can be done, but it gets complicated because we need to operate on probability functions (the Bethe measures) rather than on simple real numbers. If we just want to compute the entropy to find whether or not there exist solutions, we can take a shortcut, based on the min-sum algorithm.

Instead of computing the marginals of the distribution in Eq.~\eqref{eq:prob_SAT}, we consider the problem of minimizing the following cost (energy) function
\begin{equation}
  E(x) = \sum_{a=1}^{M} E_a(x_{\partial a}).
  \label{eq:energy}
\end{equation}
where $E_{a}(x_{\partial a}) =0$ if clause $a$ is satisfied by the assignment $x=(x_1,\dotsc,x_N)$, while $E_{a}(x_{\partial a}) =0$ otherwise. The two problems are mapped onto each other through $\psi_a(x_{\partial a}) = \mathrm{e}^{-\beta E_{a} (x_{\partial a})}$, with $\beta > 0$. The particular choice of the factor $\psi_a$ as the indicator function of clause $C_a$, Eq.~\eqref{eq:psi_indicator}, corresponds to the zero temperature limit $\beta\rightarrow \infty$.

In this formulation, the SAT-UNSAT threshold $\alpha_{\text{s}}(K)$ is identified as the value $\alpha$ above which the probability of having a configuration with ground state energy, $E(x)=0$, vanishes. 
We will estimate the ground state density with the cavity method. For this we need to adapt the message-passing rules, Eqs.~\eqref{eq:BP1}--\eqref{eq:BP2}, in two steps. First we need to compute max-marginals, rather than marginals. This is a straightforward step that consists of replacing sums with maximizations, and leads to the so-called \emph{max-product} update rules
\begin{align}
  \hat{m}^{(t)}_{a \rightarrow i}(x_i) & \cong   
  \max_{x_{\partial a \backslash i}} \biggl\{\psi_a(x_{\partial a}) \prod_{k \in \partial a\backslash i}\!m^{(t)}_{k\rightarrow a}(x_k)\biggr\},\label{eq:MP1}\\
  m^{(t + 1)}_{i \rightarrow a}(x_i) & \cong \prod_{b \in \partial i\backslash a} 
  \hat{m}^{(t)}_{b\rightarrow i}(x_i).\label{eq:MP2}
\end{align}
Second, we express these update rules in terms of the energy $E(x)$, which amounts to taking the logarithm of Eqs.~\ref{eq:MP1}--\eqref{eq:MP2}. The resulting algorithm is the so-called \emph{min-sum} algorithm:
\begin{align}
  \hat{E}^{(t)}_{a \rightarrow i}(x_i) & =   
  \min_{x_{\partial a \backslash i}} \biggl\{E_a(x_{\partial a}) + \sum_{k \in \partial a\backslash i}\!E^{(t)}_{k\rightarrow a}(x_k)\biggr\} + \hat{C}_{a \rightarrow i}^{(t)},\label{eq:MS1}\\
  E^{(t + 1)}_{i \rightarrow a}(x_i) & = \sum_{b \in \partial i\backslash a} 
  \hat{E}^{(t)}_{b\rightarrow i}(x_i) + C_{i \rightarrow a}^{(t)}.\label{eq:MS2}
\end{align}
The fixed point of these equations are known as the \emph{energetic cavity equations}. In the same way that the max-product marginals are defined up to a multiplicative constant, min-sum messages are defined up to an overall additive constant. We set the constants $C^{(t)}_{i \rightarrow a}$ and  $\hat{C}^{(t)}_{a \rightarrow i}$ so that $\min_{x_i} E_{i \rightarrow a}^{(t+1)}(x_i) = 0$ and 
$\min_{x_i} \hat{E}_{i \rightarrow a}^{(t)}(x_i) = 0$. With this arrangement, all energies are relative to the ground-state energy.

\subsection*{Warning Propagation}
The fact that the energy function, Eq.~\eqref{eq:energy}, counts the number of violated constraints allows us to simplify the min-sum updates given by Eqs.~\eqref{eq:MS1}--\eqref{eq:MS2}. It can be shown that, if messages are initialized so that $\hat{E}^{(0)}_{a \rightarrow i}$ are either 0 or 1, the subsequent values of $\hat{E}^{(t)}$ obtained from the min-sum updates will also be either 0 or 1 (see~\citep{mezard_montanari_2009}). As a consequence of this property, instead of keeping track of the variable-to-node messages $E_{i\rightarrow a}(\cdot)$, we will only bother to use the projections on $\{0, 1\}$,
\[\mathcal{E}_{i \rightarrow a} (x_i) = \min\{1, E_{i \rightarrow a} (x_i)\}.\]
The update rules become
\begin{align}
  \hat{E}^{(t)}_{a \rightarrow i}(x_i) & =   
  \min_{x_{\partial a \backslash i}} \biggl\{E_a(x_{\partial a}) + \sum_{k \in \partial a\backslash i} \mathcal{E}^{(t)}_{k\rightarrow a}(x_k)\biggr\} + \hat{C}_{a \rightarrow i}^{(t)},\label{eq:WP1}\\
  \mathcal{E}^{(t + 1)}_{i \rightarrow a}(x_i) & = \min \biggl\{1,  \sum_{b \in \partial i\backslash a} \hat{E}^{(t)}_{b\rightarrow i}(x_i) + C_{i \rightarrow a}^{(t)}\biggr\}. \label{eq:WP2}
 \end{align}
 This simplified min-sum algorithm with update equations~\eqref{eq:WP1}--\eqref{eq:WP2} is called the \emph{warning propagation} algorithm. The name stems from the interpretation of $\mathcal{E}_{i \rightarrow a}$ as a warning: $\mathcal{E}_{i \rightarrow a} = 1$ means that, according to the set of constraints $b \in \partial i \backslash a$, the $i$-th variable should not take the value $x_i$; analogously, 
$\mathcal{E}_{i \rightarrow a} = 0$ means that, according to the set of constraints $b \in \partial i \backslash i$, the $i$-th variable has green light to take the value $x_i$. The main advantage of warning propagation is that messages are are either 0 or 1, rather than distributions.

% [there is a gap here. Relate warning propagation with 1RSB cavity equations]

Because our problem involves binary variables and hard constraints, the messages of the 1RSB cavity equations are triples: $(Q_{ia}(0), Q_{ia}(1), Q_{ia}(*))$ for variable-to-function messages and $(\hat{Q}_{ai}(0), \hat{Q}_{ai}(1), \hat{Q}_{ai}(*))$ for function-to-variable messages. In the case of $K$-satisfiability, these messages can be simplified further: if $J_{ai}=0$ then $\hat{Q}_{ai}(1)$ is necessarily 0; if $J_{ai}=1$ then $\hat{Q}_{ai}(0)$ must be 0. This is because a `0' message mans that the constraint $a$ forces $x_i$ to take the value $0$ in order to minimize the system's energy. In $K$-SAT this can happen only if $J_{ai}=0$, because $x_i=0$ is the value that satisfies $a$. An analogous argument applies for the `1' message. The bottom-line is that function-to-variable messages can be parametrized by a single real number. We take this number to be $\hat{Q}_{ai}(0)$ if $J_{ai}=0$ and 
$\hat{Q}_{ai}(1)$ if $J_{ai}=1$, and denote it simply by $\hat{Q}_{ai}$.

Similarly, we can use a parametrization for the variable-to-function message $Q_{ia}(\cdot)$ that takes into account the value of $J_{ai}$. We denote by $Q_{ia}(0)$, $Q_{ia}(*)$, and $Q_{ia}(1)$ the three possible type of messages: $m(1) > m(0)=0$, $m(0) = m(1) = 0$, and $m(0) > m(1) = 0$, respectively. We then define, if $J_{ai}=0$, $Q_{ia}^{S} \equiv Q_{ia}(0)$, $Q_{ia}^{*} \equiv Q_{ia}(*)$, and $Q_{ia}^{U} \equiv Q_{ia}(1)$. Conversely, if $J_{ai}=1$, we have $Q_{ia}^{S} \equiv Q_{ia}(1)$, $Q_{ia}^{*} \equiv Q_{ia}(*)$, and $Q_{ia}^{U} \equiv Q_{ia}(0)$. The interpretation of the new defined variables is as follows
\begin{align*}
  Q_{ia}^{S} & = \Pr\bigl(x_i\ \text{is forced to satisfy $a$ by}\ b\in\mathcal{S}_{ia}\bigr),\\
  Q_{ia}^{U} & = \Pr\bigl(x_i\ \text{is forced to violate $a$ by}\ b\in\mathcal{U}_{ia}\bigr),\\
  Q_{ia}^{*} & = \Pr\bigl(x_i\ \text{is not forced by}\ b\in \mathcal{S}_{ia} \cup \mathcal{U}_{ia}\bigr),\\
  \hat{Q}_{ai} &= \Pr\bigl(x_i\ \text{is forced by clause}\ a\ \text{to satisfy  $a$}\bigr).
\end{align*}

At this point we could derive the explicit 1RSB equations in terms of the messages $Q^{S}$, $Q^{U}$, $Q^{*}$, and $\hat{Q}$. Another option is to use the above interpretation of the messages to guess the 1RSB cavity equations. Note first that clause $a$ forces variable $x_i$ to satisfy $a$ only when all the other variables involved in $a$ are forced (by some other clause) not to satisfy $a$. This can be stated as
\[ \hat{Q}_{ai} = \prod_{j \in \partial a \backslash i} Q^{U}_{ja}.\]

\begin{figure}[htbp]
  \centering
  \includegraphics{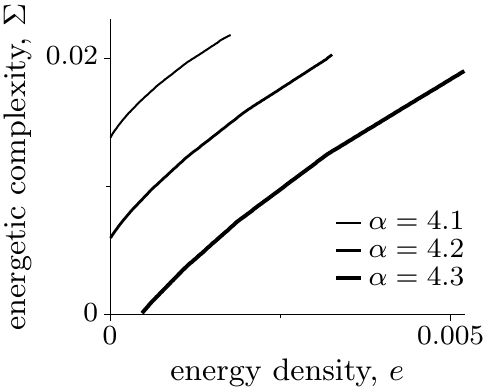}
  \caption{Energetic complexity density versus energy density for the 3-SAT problem, and for three different clause densities, indicated in the legend.}
  \label{fig:complexity}
\end{figure}

Let's define $\Omega^{S}$ and $\Omega^{U}$ as, respectively, the subset of clauses $\mathcal{S}_{ia}$ and $\mathcal{U}_{ia}$ that send a warning. For concreteness, let's pick the variable node $i$ and assume that $J_{ia}=0$ (the opposite case leads to identical equations). In that case, $\mathcal{S}_{ia}$ is the subset $b \in \partial i \backslash a$ for which $J_{ib}=0$, while $\mathcal{U}_{ia}$ is the remaining set of neighbors except $a$ for which $J_{ib}=1$. Let's also assume that the clauses $\Omega^{S} \subseteq \mathcal{S}_{ia}$ and $\Omega^{U} \subseteq \mathcal{U}_{ia}$ force the variable node $i$ to take the value $x_i$ that satisfies them. It follows that $x_i$ is forced to satisfy $a$ if $|\Omega^{S}| > |\Omega^{U}|$, and it is forced to violate $a$ if $|\Omega^{S}| < |\Omega^{U}|$; $x_i$ is not forced if $|\Omega^{S}| = |\Omega^{U}|$. The energy shift equals the number of `forcing' clauses in $\partial i \backslash a$ that are violated when $x_i$ is set to satisfy the largest number of clauses. This leads to $\min(|\Omega^{S}|, |\Omega^{U}|)$ violated clauses. The resulting 1RSB message passing algorithm, also known as Survey Propagation equations, reads
\begin{align}
  Q_{ia}^{U} & \cong \sum_{|\Omega^{U}| > |\Omega^{S}|}
  \! \mathrm{e}^{-y|\Omega^{S}|} \! \prod_{b\in \Omega^{U} \cup \Omega^{S}} \! \hat{Q}_{bi} \!\prod_{b\notin \Omega^{U} \cup \Omega^{S}} (1 - \hat{Q}_{bi}),\label{eq:SP1}\\
  Q_{ia}^{S} & \cong \sum_{|\Omega^{S}| > |\Omega^{U}|}
  \! \mathrm{e}^{-y|\Omega^{U}|} \! \prod_{b\in \Omega^{U} \cup \Omega^{S}} \! \hat{Q}_{bi} \!\prod_{b\notin \Omega^{U} \cup \Omega^{S}} (1 - \hat{Q}_{bi}),\label{eq:SP2}\\
  Q^{*}_{ia} &\cong \sum_{|\Omega^{U}| = |\Omega^{S}|}
  \! \mathrm{e}^{-y|\Omega^{U}|} \! \prod_{b\in \Omega^{U} \cup \Omega^{S}} \! \hat{Q}_{bi} \!\prod_{b\notin \Omega^{U} \cup \Omega^{S}} (1 - \hat{Q}_{bi}).\label{eq:SP3}
\end{align}
The overall normalization is fixed by the condition $Q_{ia}^{U} + Q_{ia}^{S} +Q_{ia}^{*} = 1$. These equations are not much more complicated to solve than those for BP. Like in the BP equations, we can use Eqs.~\eqref{eq:SP1}--\eqref{eq:SP3} to find the fixed point of the messages $\{\hat{Q}_{ai}, Q_{ia}\}$ for a given instance, or, rather, we can do statistical analysis. In the latter case, we can compute with population dynamics the probabilities $P(\hat{Q}_{ai})$ and $P(Q^{U}_{ia}, Q^{S}_{ia}, Q^{*}_{ia})$. We can then compute the Bethe-free energy, and then the Legendre transform of the resulting formula, from which we obtain the complexity as a function of the energy. We get Figure~\ref{fig:complexity}
From the figure we see that $\alpha = 4.3$ we get a certain number of contradictions (given by the finite energy at $\Sigma=0$, i.e., the intersection with the abscissa). The number of contradictions decreases as we reduce $\alpha$, until contradictions vanish. This happens when the value of $\alpha$ is such that the curve crosses the origin of the $\Sigma$ vs energy curve, which is approximately $\alpha \approx 4.2667$. This is the prediction for the SAT-UNSAT threshold. An analogous derivation for the 4-SAT problem leads to the estimate $\alpha \approx 9.667$. 

% We can use the same method singla instance, using decimation. Look at the node receiving more warnings, fix it. We can compute the complexity curve.

%% Create Chapter Biblio. %%%%%%%%%%%%
%% Uncomment these lines to add in your bibliography
\bibliographystyle{OUPnamed}
\addcontentsline{toc}{section}{References}
\bibliography{references}

\thebibliography{0}

\bibitem[\protect\citeauthoryear{Almeida and Thouless}{Almeida and
  Thouless}{1978}]{de1978stability}
Almeida, JRL~De and Thouless, David~J. (1978).
\newblock Stability of the {S}herrington-{K}irkpatrick solution of a spin glass
  model.
\newblock {\em J. Phys. A\/},~{\bf 11}(5), 983.

\bibitem[\protect\citeauthoryear{Edwards and Anderson}{Edwards and
  Anderson}{1975}]{edwards1975theory}
Edwards, Samuel~Frederick and Anderson, Phil~W (1975).
\newblock Theory of spin glasses.
\newblock {\em J. Phys. F\/},~{\bf 5}(5), 965.

\bibitem[\protect\citeauthoryear{M\'{e}zard and Montanari}{M\'{e}zard and
  Montanari}{2009}]{mezard_montanari_2009}
M\'{e}zard, Marc and Montanari, Andrea (2009).
\newblock {\em {Information, Physics, and Computation}}.
\newblock Oxford University Press.

\bibitem[\protect\citeauthoryear{M\'{e}zard and Mora}{M\'{e}zard and
  Mora}{2009}]{mezard_mora_jphysiolp2009}
M\'{e}zard, Marc and Mora, Thierry (2009).
\newblock Constraint satisfaction problems and neural networks: {A} statistical
  physics perspective.
\newblock {\em J. Physiol.-Paris\/},~{\bf 103}(1), 107--113.

\bibitem[\protect\citeauthoryear{M\'{e}zard and Parisi}{M\'{e}zard and
  Parisi}{2001}]{Mezard2001}
M\'{e}zard, Marc and Parisi, Giorgio (2001).
\newblock {The Bethe lattice spin glass revisited}.
\newblock {\em Euro. Phys. J. B\/},~{\bf 233}, 217--233.

\bibitem[\protect\citeauthoryear{M\'{e}zard and Parisi}{M\'{e}zard and
  Parisi}{2003}]{Mezard2003}
M\'{e}zard, Marc and Parisi, Giorgio (2003).
\newblock {The cavity method at zero temperature}.
\newblock {\em J. Stat. Phys.\/},~{\bf 111}(April).

\bibitem[\protect\citeauthoryear{M\'{e}zard, Parisi and Virasoro}{M\'{e}zard
  {\em et~al.}}{1986}]{mezard1986sk}
M\'{e}zard, Marc, Parisi, Giorgio, and Virasoro, Miguel~{\'A}ngel (1986).
\newblock {SK} model: The replica solution without replicas.
\newblock {\em Europhys. Lett\/},~{\bf 1}(2), 77--82.

\bibitem[\protect\citeauthoryear{Parisi}{Parisi}{1979}]{parisi1979infinite}
Parisi, Giorgio (1979).
\newblock Infinite number of order parameters for spin-glasses.
\newblock {\em Phys. Rev. Lett.\/},~{\bf 43}(23), 1754.

\bibitem[\protect\citeauthoryear{Parisi}{Parisi}{1980}]{parisi1980order}
Parisi, Giorgio (1980).
\newblock The order parameter for spin glasses: {A} function on the interval
  0-1.
\newblock {\em J. Phys. A\/},~{\bf 13}(3), 1101.

\bibitem[\protect\citeauthoryear{{Sherrington} and {Kirkpatrick}}{{Sherrington}
  and {Kirkpatrick}}{1975}]{sk1975prl}
{Sherrington}, D. and {Kirkpatrick}, S. (1975, December).
\newblock {Solvable Model of a Spin-Glass}.
\newblock {\em Phys. Rev. Lett.\/},~{\bf 35}, 1792--1796.

\bibitem[\protect\citeauthoryear{Thouless, Anderson and Palmer}{Thouless {\em
  et~al.}}{1977}]{thouless1977solution}
Thouless, DJ, Anderson, PW, and Palmer, RG (1977).
\newblock Solution of 'solvable model of a spin glass'.
\newblock {\em Philos. Mag.\/},~{\bf 35}(3), 593--601.

\bibitem[\protect\citeauthoryear{Touchette}{Touchette}{2009}]{Touchette2009}
Touchette, Hugo (2009).
\newblock {The large deviation approach to statistical mechanics}.
\newblock {\em Phys. Rep.\/},~{\bf 478}(1-3), 1--69.

\endthebibliography
\end{document}